\documentclass[letterpaper,11pt]{article}

\usepackage{jheppub, tikz, physics, enumitem, tikz-feynman, indentfirst, color, bbold, comment, parskip}

\usepackage[mathscr]{eucal}

\usepackage[linkcolor=blue, colorlinks=true, citecolor=blue, urlcolor=blue]{hyperref}

\def\equationautorefname~#1\null{eq. (#1)\null}

\tikzfeynmanset{compat=1.1.0}

\newcommand{\n}{d}

\newcommand{\be}{\begin{equation}}
\newcommand{\ee}{\end{equation}}

\usetikzlibrary{intersections, shapes.arrows, shapes.misc, decorations.pathreplacing, decorations.pathmorphing, angles, quotes, calc, patterns, arrows.meta, fadings, tikzmark}

\tikzset{cross/.style={cross out, draw=black, minimum size=2*(#1-\pgflinewidth), inner sep=0pt, outer sep=0pt}, cross/.default={1pt}}

\allowdisplaybreaks

\title{\boldmath Hadamard tails from flat space perturbation theory}

\author[a]{Ameya Chavda,}
\author[a]{Alberto Nicolis,}
\author[b]{Alessandro Podo,}
\author[a, c]{John Staunton}

\affiliation[a]{Center for Theoretical Physics and Department of Physics, Columbia University, \\
New York, NY 10027, USA}

\affiliation[b]{Institut des Hautes \'Etudes Scientifiques,\\
91440 Bures-sur-Yvette, France}

\affiliation[c]{Center for Particle Cosmology, Department of Physics and Astronomy,\\ University of Pennsylvania, 209 South 33rd St, Philadelphia, PA 19104, USA}

\emailAdd{ameya.chavda@columbia.edu}
\emailAdd{a.nicolis@columbia.edu}
\emailAdd{podo@ihes.fr}
\emailAdd{jstaunto@sas.upenn.edu}

\abstract{The short-distance singular structure of the two-point function of a free scalar field in curved spacetime has a universal behavior that characterizes well-behaved states (called Hadamard states). This includes a non-analytic term proportional to the Ricci scalar curvature known as the Hadamard tail. This is usually derived by solving a differential equation for the Green's function of a Klein-Gordon field in curved spacetime. We present an alternative derivation which leverages the equivalence principle and makes use of perturbative field theory methods. This allows for the computation of the short-distance singular behavior of correlators of QFTs in curved space. As examples, we apply these ideas to the two-point function of massless spin-1 and spin-2 fields in an arbitrary curved background, in general dimension.  We also extend these results to the two-point correlators of scalar primary operators of an arbitrary Conformal Field Theory coupled to an arbitrary gravitational background.}

\begin{document}
\maketitle
\flushbottom

\section{Introduction}
\label{sec:intro}
\textbf{Background.} Quantum field theory (QFT) in curved spacetime is a vast subject with many applications but, unfortunately, limited capability for exact or general results. For example, many results from QFT in curved spacetime have only been obtained for highly symmetric spacetimes or for free fields, see for instance Ref.~\cite{Hollands:2014eia} for a review using the algebraic approach.\footnote{There are some exceptions: for instance, central charges associated with trace anomalies of conformal field theories in a curved background can often be determined exactly in superconformal field theories, see e.g. Refs.~\cite{Shapere:2008zf,Martone:2020nsy,Rastelli:2023sfk} and references therein.} It is thus important to develop systematic approximation schemes that are applicable to generic QFTs in arbitrary curved backgrounds.

One such scheme leverages the equivalence principle: for length-scales much shorter than the curvature radius, all regular metrics are approximately flat. This was used to compute quantum effects in a cosmological background in Ref.~\cite{Komissarov:2022gax}. Thus, in principle, one can use standard perturbation theory for QFT in \emph{flat} spacetime to compute the effects of curvature, treating the background metric's deviation from flatness as a small perturbation. Similar ideas have also been explored in the context of black hole physics, in particular in the effective field theory approach to classical self-force computations \cite{Kosmopoulos:2023bwc,Cheung:2023lnj} as well as in novel derivations of Hawking radiation \cite{Aoude:2024sve}.

In this paper, we apply this idea to the short-distance behavior of correlation functions. For a free and minimally coupled scalar field, $\phi$, in four dimensions, it is well known that the $T$-ordered correlation function takes the form \cite{DeWitt:1960fc, Parker:2009uva}:
\begin{equation}\label{eq:Hadamard}
G_F\left(x_1, x_2\right) \equiv \mel{0}{T\phi\left(x_1\right)\phi\left(x_2\right)}{0} = \frac{1}{4\pi^2} \left( \frac{u}{s^2 + i\epsilon} + v \log\left(s^2 + i \epsilon\right)+w \right),
\end{equation}
where $s$ is the geodesic distance between $x_1$ and $x_2$, and $u$, $v$, and $w$ are smooth functions of $x_1$ and $x_2$ that reduce to $u=1$, $v=0$ and $w=0$ in the flat space limit. The term $v \log\left(s^2 + i \epsilon\right)$ is known as the \emph{Hadamard tail}, and the value of $v$ in the coincident limit $x_1 \to x_2$ is a universal prediction of QFT in curved spacetime, which in our conventions takes the form
\begin{equation}
\label{eq:Limit of v}
\lim\limits_{x_1\to x_2} v = -\frac{R}{24} \; ,
\end{equation}
with $R$ being the local value of the Ricci scalar. By universal, we mean that this coefficient is fixed by the local geometry and is state-independent.

To gain a better physical understanding of the above result, it is useful to look at the retarded correlator, which can be expressed in terms of $G_F$ as 
\begin{equation}
G_R= 2i \theta(t_1-t_2) \, \operatorname{Im}G_F,
\end{equation}
so that
\begin{equation}
\label{eq:GR}
G_R\left(x_1, x_2\right) = - \frac{i}{2\pi} \theta(t_1- t_2) \left(u \,\delta(s^2) -v \theta(- s^2)\right).
\end{equation}
We see that in flat space, $v=0$, $u=1$, the retarded correlator of a massless scalar field is supported on the future lightcone, whereas in curved spacetime there is an additional contribution that is supported \emph{inside} the lightcone ($s^2<0$), as expected generally as shown by investigations of microcausality in Lorentz-breaking states \cite{Dubovsky:2007ac, Hui:2025aja}. This contribution is proportional to $v$ and hence is a consequence of the presence of the Hadamard tail.

\textbf{Motivation.} Due to its universal nature, the Hadamard form of two-point functions has appeared in many seemingly unrelated calculations and therefore merits further investigation. In particular, it  is used to produce the nonlocal, history-dependent part of the self-force acting on a compact object \cite{DeWitt:1960fc, Mino:1996nk, Quinn:1996am, Detweiler:2002mi}. Moreover, subtraction of the local universal singular part underlies covariant point-splitting definitions of the renoralized stress-energy tensor and therefore enters semiclassical backreaction and the calculation of trace anomalies \cite{Christensen:1976vb, Wald:1977up, Decanini:2005eg}. The tail has also been computed perturbatively in weakly curved spacetimes \cite{Pfenning:2000zf} and for gravitational-wave propagation and scattering in perturbed backgrounds \cite{Chu:2011ip, Kessler:2025idf}.

Knowledge of the non-analytic and short-distance structure of the graviton propagator in a general curved background and in general dimensions can also prove very useful in the context of gravitational self-force computations~\cite{Pound:2015tma}, in particular for analytic approaches~\cite{Kavanagh:2026ivj}, where it could open the way to dimensionally regularization. Our final results, in particular the graviton propagator, hold to leading order in the curvature, but are exact and do not require any PM or PN expansion of the background metric.

Finally, the Hadamard form of correlation functions underlies several axiomatic efforts to formalize quantum field theory on curved spacetimes \cite{Radzikowski:1996pa, Brunetti:1995rf, Hollands:2001nf, Hollands:2001fb, Hollands:2009bke, Hollands:2023txn}. In this framework, the structure of short-distance singularities of correlation functions in a Hadamard state is assumed to be restricted by the so-called ``microlocal spectral conditions,'' one of which is that the wavefront set of the two-point function should take a Hadamard form, as in Eq.~\eqref{eq:Hadamard} \cite{Radzikowski:1996pa}. Roughly speaking, the wavefront set of a distribution is a sharper description of its singular support by specifying, for each point at which the distribution is singular, the momentum space directions along which its Fourier transform fails to fall off rapidly.

\textbf{Summary of Results.} Our first, modest goal in this paper is to rederive Eqs.~\eqref{eq:Hadamard} and~\eqref{eq:Limit of v} from QFT perturbation theory.  The method we  develop will then lend itself to a number of generalizations. In particular, we will compute the universal singular parts of correlation functions of spin 1 and spin 2 fields, and scalar operators in interacting QFTs coupled to a gravitational background. As a consistency check, we will also use this approach to rederive the conformal anomaly of 2D Conformal Field Theories (CFTs) in a gravitational background. In a companion paper, some of us apply these methods to the study of the singular structure of higher-point correlation functions in interacting quantum field theories in curved spacetime. 

A few qualifications are in order. First, in the first part of the paper we treat the curved spacetime metric as an external, non-dynamical background. It is in principle straightforward to treat its quantum fluctuations as a dynamical field, like one does, for example, for the tensor correlation functions in inflationary cosmology. We do so in sect.~\ref{spin 2}, but for simplicity we restrict to pure-gravity cases.

Second, in a generic curved spacetime with no global timelike Killing vector, there is no unambiguous characterization of ``the" vacuum. For this reason, one usually considers generic states that, at the level of correlation functions, reduce to the standard Poincar\'e-invariant vacuum at short enough distances, where the effects of curvature can be neglected in first approximation. These are called ``Hadamard states" \cite{Hollands:2014eia}, and the result in Eq.~\eqref{eq:Hadamard} holds for those. For example, the Bunch-Davies vacuum in de Sitter space is a Hadamard state. On the other hand, there is a family of de Sitter-invariant ``vacua", known as $\alpha$-vacua, that are not Hadamard. Correlators in these states are known to have antipodal singularities; moreover, in the coincident limit, they have a different behavior from that of correlators in the Minkowski vacuum. The only exception is the Bunch-Davies vacuum itself, which corresponds to $\alpha \to -\infty$ \cite{Kaloper:2002cs}.

Notice that all states that we usually consider in flat space QFT, including arbitrary multi-particle asymptotic states, are Hadamard states: as long as a state is not infinitely excited in the UV, its correlation functions reduce to the vacuum ones at very short distances. There can be subleading singularities, such as the log in Eq.~\eqref{eq:Hadamard}, but the leading singularity is the same as in the vacuum. This is to say that the restriction to Hadamard states is extremely reasonable and generous. In our approach, where we start with the flat spacetime vacuum and we apply perturbation theory, this condition will be implemented automatically.

\textbf{Comparison to Previous Literature.} The classic result in Eq.~\eqref{eq:Hadamard} has been historically derived by taking advantage of the fact that, for a free field, two-point correlators are Green's functions for the Laplace-Beltrami operator in the curved background, a classical mathematical problem already analyzed by Hadamard \cite{Hadamard:1923}. This method was applied  in four dimensions for spin-0 and spin-1 fields in Ref.~\cite{DeWitt:1960fc} and later extended for four-dimensional spin-2 fields in Ref.~\cite{Allen:1987bn}. The same equations were also studied in a local momentum-space representation, introduced by Bunch and Parker \cite{Bunch:1979uk} and relying on the use of Riemann normal coordinates: for scalar fields in Ref.~\cite{Parker:2009uva} and spin-1 fields in Ref.~\cite{Toms:2014tia} in general dimensions. Recursion relations for the spin-2 Hadamard coefficients were then given in general dimensions in Ref.~\cite{Frob:2017gez}. In this paper, we give explicit expressions for the leading coefficient.

Each of these methods provides a generally covariant approach for the computation of operator product expansion (OPE) coefficients, but  explicitly extracting the local coefficients can be technically demanding. 
In this work, we develop a complementary diagrammatic implementation. We represent weak curvature by soft metric insertions in ordinary flat-space correlators, and obtain the short-distance expansion through a systematic hard-soft momentum expansion. The computation reduces to a series of one-loop integrals, which can then be applied verbatim to interacting theories, where the approach of Ref.~\cite{DeWitt:1960fc} cannot be directly applied. In our approach, we postpone the closed form expression of the generally covariant result until the end of the calculation, so that intermediate steps in the calculation are simplified. We mention here that diagrammatic approaches have been used also in the derivation of conformal anomalies, for example in Refs.~\cite{Godazgar:2018boc, Casarin:2018odz}. 

Beyond free or weakly coupled theories, one can study the gravitational corrections to correlators in strongly interacting QFTs, such as conformal field theories coupled to a background metric. One method for doing so has been presented in Ref.~\cite{Konechny:2026bqg},\footnote{Ref.~\cite{Konechny:2026bqg} appeared four months after the first version of this work was shared on arXiv.} which computed generally covariant OPEs for a $d$-dimensional Euclidean CFT. There, they focused on conformally flat metrics, but argued that their results would apply more generally. In the aforementioned companion paper, we compute OPE coefficients for scalar primary operators, but for a general curved background. Our results agree with those of Ref.~\cite{Konechny:2026bqg} where the two overlap.

\textbf{Notation and Conventions.} Throughout this paper, we will work in $d$ spacetime dimensions, unless otherwise specified, and use the mostly plus signature $\left(-, +, \dots, +\right)$. General Relativity (GR) conventions will closely follow those of Ref.~\cite{Wald:1984rg}. In particular, the Riemann tensor is
\begin{equation}
R_{\mu\nu\rho} {}^{\sigma} = \partial_\nu \Gamma^\sigma_{\hphantom{\sigma}\mu\rho} - \partial_\mu \Gamma^\sigma_{\hphantom{\sigma}\nu\rho} + \Gamma^\lambda_{\hphantom{\lambda}\mu\rho}\Gamma^\sigma_{\hphantom{\sigma}\lambda\nu} - \Gamma^\lambda_{\hphantom{\lambda}\nu\rho}\Gamma^\sigma_{\hphantom{\sigma}\lambda\mu}.
\end{equation}
Moreover, the Ricci tensor is $R_{\mu\nu} = R_{\mu\rho\nu}^{\hphantom{\mu\rho\nu}\rho}$ and the Ricci scalar is $R = g^{\mu\nu} R_{\mu\nu}$. Our Fourier transform conventions are as follows:
\begin{equation}
\tilde{f}(k) = \int \! f(x)\,  e^{-ik \cdot x}\, \dd[\n]{x} \qquad \text{and} \qquad f(x) = \int \! \tilde{f}(k) \, e^{ik\cdot x} \, \frac{\dd[\n]{k}}{\left(2\pi\right)^\n}.
\end{equation}
To simplify the notation we will drop the tilde from Fourier transformed quantities, relying on the argument (position or momenta) to distinguish between $f(x)$ and $f(k)\equiv \tilde{f}(k)$.
We will also often use the following abbreviation of momentum space integrals:
\begin{equation}
\int_{k_1, \dots,\, k_l} \equiv \int \frac{\dd[\n]{k_1}}{\left(2\pi\right)^\n} \dots \frac{\dd[\n]{k_l}}{\left(2\pi\right)^\n}.
\end{equation}
Finally, we will use  $x^2 \equiv \eta_{\mu\nu} x^\mu x^\nu$ for squared distances as measured by the flat-space metric, $s^2$ for squared geodesic distances as measured by the curved metric, $ds^2 = g_{\mu\nu} \dd{x}^\mu \dd{x}^\nu$, and we will raise and lower indices using $\eta_{\mu\nu}$ only.

\section{The general framework for correlation functions}

Consider a generic QFT in a curved, but non-dynamical, spacetime. The action $S = S[\phi, g]$ is a functional of the matter fields, collectively denoted by $\phi$, and of the spacetime metric $g_{\mu\nu}$.  
For studying correlation functions at short distances compared to the radius of curvature, one can choose coordinates in which the metric is locally approximately flat: $g_{\mu\nu} (x)= \eta_{\mu\nu} + h_{\mu\nu}(x)$, with $h_{\mu\nu}\ll 1$. One can then expand the action in powers of $h_{\mu\nu}$, and use perturbation theory treating $h_{\mu\nu}$ as an external source. To first order,
\be
S [\phi, g] \simeq S_{0}[\phi] + \frac{1}{2}\int h_{\mu\nu}\left(x\right) T^{\mu\nu} \left(x\right)\dd[\n]{x},
\ee
where $S_{0}[\phi] \equiv S [\phi, \eta]$ is the flat-space action for the matter fields, and $T^{\mu\nu}$ is their gravitational stress-energy tensor evaluated in flat space,
\begin{equation}
\label{eq:TcanGrav}
T_{\mu\nu} (x) \equiv -\dfrac{2}{\sqrt{-g}}\dfrac{\delta S[\phi, g]}{\delta g^{\mu\nu}} \bigg|_{g=\eta} \; .
\end{equation}

Now, suppose one wants to compute, in curved spacetime, the short-distance behavior of the $T$-ordered correlation function of two generic operators, $O_1$ and $O_2$. By applying standard flat space perturbation theory, to first order in $h_{\mu\nu}$ one simply has
\begin{equation}
\begin{aligned}
\expval{ T \big(O_1(x_1) O_2(x_2) \big)}_g & \simeq \langle T \big(O_1(x_1) O_2(x_2) \big) \rangle_{0} \label{eq:general}\\ 
& \quad + \frac{i}{2} \int \dd[d]{y}  \, \langle T \big(O_1(x_1) T^{\mu\nu} (y) O_2(x_2) \big)\rangle_{0} \, h_{\mu\nu}(y) \;, \end{aligned}
\end{equation}
where the correlation functions on the right-hand side, $\langle \dots \rangle_0$, are evaluated in flat spacetime, i.e. from the flat spacetime action $S_0[\phi]$.

Recall that we are expanding in $h_{\mu\nu}$ because of the equivalence principle: at short distances, deviations from flatness are small. We thus see that the leading short-distance correction to a two-point function only depends on a three-point function, that of the operators of interest together with the stress-energy tensor. For a free or perturbative theory, this can be computed systematically; for a  CFT, as long as $O_1$ and $O_2$ are primary operators, it takes a universal form (see e.g. Ref.~\cite{Simmons-Duffin:2016gjk}). 

In the explicit computations that we perform below, some other features will emerge. First, in the case of a perturbative diagrammatic approach, it is convenient to work in momentum space. One can identify a hard momentum $p$, associated with $x_1-x_2$, and a soft one, $q$, carried by $h_{\mu\nu}$. Working at short distances compared to the radius of curvature corresponds to the limit $q \ll p$. This is what we will do in the case of the (gravitationally coupled) free scalar.\footnote{The momentum-space OPE can fail when the operator with small momentum has a large scaling dimension, because of divergences in the resulting integrals~\cite{Dymarsky:2014zja}. It always works, on the other hand, for correlators of ``elementary'' fields (weakly coupled fields with an ordinary free action plus interactions), to all orders in perturbation theory~\cite{Collins:1984xc}.} Then, to first order in $h_{\mu\nu}$, we know that the first gauge-invariant objects we can construct are of order $\partial^2 h$, and so in momentum space we will have to work to second order in $q/p \ll 1$. We will later extend our results to interacting CFTs in Section~\ref{s:Conformal Primaries}, working directly in position space and taking advantage of conformal invariance. In that case we will directly expand the metric to second order in derivatives.

The conversion between the flat space squared distance $(x_1 -x_2)^2$ and the curved space  geodesic squared distance $s^2$ involves $h_{\mu\nu}$. And so, rewriting our results in a gauge-invariant, GR-friendly format will require some care.  In fact, there will be additional subtleties associated with the fact that a two-point function of covariant operators should transform as a {\em bi}-tensor under diffeomorphisms. We review the properties and expansions of common bi-tensors in Appendix~\ref{a:GeometricInvariants}.

\section{Why (and when) it works}\label{s:why}
The explicit computations that we perform below are admittedly quite dry, and they obscure some important conceptual and technical aspects of our approach. We thus want to briefly address those here. 

At face value, Eq.~\eqref{eq:general} is only predictive if one knows $h_{\mu\nu}$ everywhere, and if one is able to perform the integral. This goes against the spirit of our approach: to figure out the effects of gravity locally, at short length-scales, one should only need to know the local values of some geometric quantities. In particular, one expects that the singular behavior of correlation functions at short distance should be determined by the local geometry, with no reference to what happens at large distances.

Notice, however, that at some order in the short-distance expansion, one does expect the contributions due to a nontrivial geometry to become sensitive to some global properties. For instance, for a massless free field on a cylinder of radius $R$ (which has a {\em flat} geometry), the two-point function is a sum over images, which in the short-distance limit ($x \ll R$) features a relative correction of order $(x/R)$ to some power, compared to the infinite-space two-point function.

Perhaps related to this, there is also the ambiguity of which state we are in. As we briefly reviewed in Section~\ref{sec:intro}, in a generic geometry there is no preferred state playing the role of the vacuum. The best we can hope for is to restrict to a large class of well-behaved states---the Hadamard ones---and look for general statements that apply to those. 
But correlation functions are state-dependent at large enough distances compared to the scales that characterize the state.
For example, in flat space, a single particle state of spatial momentum $p$ and the vacuum differ significantly at the level of correlation functions at distances of order $1/p$ and beyond. 

Finally, the equivalence principle guarantees that deviations from flatness are small at short distances, but at large enough distances the expansion in powers of $h_{\mu\nu}$ in general breaks down. So, our truncated expansion at first order in $h_{\mu\nu}$ cannot be justified for large distances, except in special cases, such as that of a small-amplitude gravitational wave.

It is thus crucial to understand what kind of information can be reliably derived from Eq.~\eqref{eq:general}.
We expect that Eq.~\eqref{eq:general} allows us to compute the short-distance singular behavior of two point-functions, including all subleading singularities, but is not predictive, in general, for corrections that are analytic in $\Delta x^\mu \equiv (x_1-x_2)^\mu$.  \emph{Mutatis mutandis}, this sounds very much like the familiar statement in quantum field theory that a low-energy effective field theory is all that is needed to compute the non-analytic terms in observables at low momenta and energies, while analytic pieces in general depend on the high-energy dynamics. There is one important difference, though: usually, we work at low momenta/long distances, and we are interested in non-analycities in $p^\mu$, for $p\to0$. Here, instead, we are working at {\em short} distances, and we are interested in non-analyticities in $\Delta x^\mu$, for $\Delta x\to 0$. 

Still, the technical reason for why the two statements are both valid is pretty much the same. To see this, it is useful to consider our perturbation theory in momentum space. As we will see, at least at the order at which we will be working, the $\Delta x^\mu$ dependence of our two-point functions will come from inverse Fourier transforms of the form\footnote{We consider a massless theory, for simplicity and also because in a massive theory one expects correlation functions to be exponentially suppressed at large distances. In that case, our analysis would  be  simpler---the subtleties that we address in this section are much more relevant for the massless case.}
\be
I_\alpha\left(\Delta x\right) = \int \frac{\dd[\n]{k}}{\left(2\pi\right)^\n}
 \frac{1}{k^{2{\alpha}}} \, e^{i k\cdot \Delta x}  \; ,
\ee
which in this section, for simplicity, we are considering in Euclidean signature (both for $\Delta x^\mu$ and~$k^\mu$).

Let's consider this integral at finite Euclidean separations, $\Delta x^2>0$.
Then, the integral converges in the UV, at large $k$, because the oscillatory exponential acts effectively as a UV cutoff of order $k \sim 1/\Delta x$. However, the integral diverges in the IR, for $k \to 0$, whenever
\be
\alpha - d/2 \ge 0 \; .
\ee
The fact that $I_\alpha(\Delta)$ can  be IR-divergent tells us that, in general, our two-point function will depend strongly on how the divergence is cutoff in the IR, that is, on what the geometry looks like far from the point around which we are expanding. 

As in the case of standard UV divergences in QFT, however, we can make the integral convergent by differentiating it enough times with respect to the right variable \cite{Weinberg:1995mt}. Before we do so, it is convenient to rewrite the integrand in Schwinger parametrization,
\be
\frac{1}{k^{2\alpha}} = \frac{1}{\Gamma(\alpha)} \int_0^\infty \! ds \, s^{\alpha-1} e^{-s \,k^2 } \; .
\ee
Then, the integral in $k^\mu$ is a convergent Gaussian integral, and we obtain
\be \label{eq:Schwinger}
I_\alpha (\Delta x^2) = \frac{1}{(4 \pi)^{d/2} \Gamma(\alpha)} \int_0 ^\infty \! ds \, s^{\alpha-d/2-1} e^{-\frac{ \Delta x^2}{4s}}  \; . 
\ee
The advantage of this rewriting is that now $I_\alpha$ is explicitly a function of $\Delta x^2$. Notice that, for positive $\Delta x^2$, the integral in $s$ is convergent at $s =0$, but diverges for $s \to \infty$ for $\alpha-d/2 \ge 0$. This is the same conditions as for the IR divergence that we discussed above, and so, in this parametrization, $s \to 0$ corresponds to the UV, and $s \to \infty$ to the IR.

Now, the crucial observation is that if we differentiate enough times with respect to $\Delta x^2$, we get a convergent integral:
\be
 I^{(N)}_\alpha( \Delta x^2)  \propto \int_0 ^\infty \! ds \, s^{\alpha-N-d/2-1} e^{-\frac{ \Delta x^2}{4s}}  \; ,
\ee
which is finite in the IR as long as 
\be
N > \alpha-d/2 \; .
\ee
Such an integral is in general a non-analytic function of $\Delta x^2$, because the exponential factor in Eq.~\eqref{eq:Schwinger} kills a possible UV divergence at $s=0$ only for nonzero $\Delta x^2$. So, we do expect singularities at $\Delta x = 0$---and that is in fact the point of our paper---but, at least after taking enough derivatives with respect to~$\Delta x^2$, we see that they appear with IR-finite coefficients. 

Then, we complete the argument in the usual way: we can reconstruct the original integral by integrating $N$ times in $\Delta x^2$ and introducing $N$ integration constants:
\begin{equation}
\begin{split}
I_\alpha (\Delta x^2) & = \int d(\Delta x^2) \cdots \int d(\Delta x^2)  \, I^{(N)}_\alpha( \Delta x^2) \\
& + A_1 + A_2 \Delta x^2 + \cdots + A_N (\Delta x^2)^{N-1} \; .
\end{split}
\end{equation}
The indefinite integrals in the first line yield the $N$-th order primitive of $I^{(N)}$, which is in general non-analytic in $\Delta x^2$, but finite for nonzero, positive $\Delta x^2$.\footnote{For example, we could compute it by choosing a nonzero, positive lower extremum for all the integrals, so that by integrating we never cross the non-analyticity point $\Delta x^2 = 0$.} The integration constants on the second line, $A_1, \dots, A_N$ are in general IR divergent, but the crucial observation is that they multiply terms that are {\it analytic} in $\Delta x^2$.

In conclusion, assuming IR divergences are a reliable proxy for our results' sensitivity to the physics and geometry at very large distances from our expansion point, we see that our framework and perturbative techniques allow us to compute all non-analytic contributions to two-point functions, that is, singularities for $\Delta x^2 \to 0$. Analytic terms, on the contrary, are sensitive to the details of what happens at very large distances. With this understanding, from now on we will write expressions for correlators up to terms that are analytic  in the coincident limit. Moreover, we will always work at leading order in $h_{\mu \nu}$ and up to second order in derivatives.

\section{Scalar two-point functions: the free field} \label{s:spin0}

We start by developing the blueprint described in the previous two sections to the computation of two point functions of a free massless scalar, $\phi$, in general spacetime dimensions $d$, with minimal and non-minimal coupling. Where appropriate, we compare our answers with results in the literature. 

Consider the action of a minimally coupled free scalar, 
\begin{equation} \label{eq:freeAction}
S[\phi,g] = -\frac{1}{2}\int g^{\mu\nu}\left(x\right)\partial_\mu \phi\left(x\right)\partial_\nu \phi\left(x\right)\sqrt{-g}\dd[\n]{x} \; .
\end{equation}
The flat-space gravitational stress-energy tensor is
\begin{equation}
\label{eq:Tcan}
T_{\mu\nu}= \partial_\mu \phi \partial_\nu \phi - \frac{1}{2} \eta_{\mu\nu} \partial_\lambda \phi \partial^\lambda \phi,
\end{equation}
which in this case  coincides with the canonical stress-energy tensor (i.e. the one derived from the Noether procedure for spacetime translations).

We now apply Eq.~\eqref{eq:general} to the $\phi\phi$ $T$-ordered correlation function,
\be
G_F\left(x_1, x_2\right)=\expval{T\phi\left(x_1\right)\phi\left(x_2\right)}_{g} \; .
\ee
As mentioned above, in a perturbative diagrammatic computation it is more convenient to work in momentum space.

The Feynman rules are easily derived.
The propagator for $\phi$ is the flat space one
\begin{equation}
\label{eq:propagator}
    \begin{gathered}
    \vspace{-0.5cm}
        \feynmandiagram[horizontal=a to b, layered layout]
        {a -- [plain, momentum'={$k$}] b};
    \end{gathered} = -\frac{i}{k^2-i\epsilon}.
\end{equation}
The interaction between~$\phi$ and the background metric perturbation $h_{\mu\nu}$ is given by the following vertex:
\begin{equation}
\label{eq:hinteractionvertex}
    \begin{gathered}
    \vspace{-0.5cm}
        \feynmandiagram[horizontal=a to c, layered layout]
        {a -- [plain, momentum'={$k$}] b[crossed dot, minimum size=3mm] -- [plain, momentum'={$k'$}] c};
    \end{gathered} = \frac{i}{2}\left(k^\mu k'^\nu + k'^\mu k^\nu\right)\bar{h}_{\mu\nu}\left(k' - k\right),
\end{equation}
where 
\begin{equation}
\label{eq:hbar}
\bar{h}_{\mu\nu} = h_{\mu\nu} - \frac{1}{2}\eta_{\mu\nu}h \qquad {\rm and } \qquad 
 h \equiv  \eta^{\mu\nu} h_{\mu\nu}.
\end{equation}
With these, we can write a diagrammatic expansion for the two-point correlation function in powers of $h$, represented by the crossed-circle interaction vertex. To first order, we have
\begin{equation}
    G_F(x_1,x_2)\equiv \expval{T\phi\left(x_1\right)\phi\left(x_2\right)}_{g} = \begin{gathered}
    \vspace{0.3cm}
    \feynmandiagram[horizontal = a to c]{a[dot, label=above:{$x_1$}] -- c[dot, label=above:{$x_2$}]};
    \end{gathered} + \begin{gathered}
    \vspace{-0.45cm}
    \feynmandiagram[horizontal = a to c, layered layout]{a[dot, label=above:{$x_1$}] -- [plain, momentum'={$k$}] b [crossed dot, minimum size=3mm] -- [plain, momentum'={$k'$}] c[dot, label=above:{$x_2$}]};
    \end{gathered} + \dots.
\end{equation}
The first diagram is the flat space propagator, while the second diagram corresponds to the first-order gravitational correction:
\begin{equation}
G_F(x_1,x_2) = G_F^{(0)}\left(\Delta x\right) + G_F^{(1)}\left(\Delta x, x_0\right) + \dots,
\end{equation}
where the superscript denotes the order in the weak field expansion $h_{\mu\nu}\ll 1$.
Note that $G_F^{(0)}$ depends only on the difference $\Delta x = x_1 - x_2$ due to translation invariance. On the other hand, $G_F^{(1)}$ depends on both $\Delta x$ and $x_0$, where $x_0$ is the point around which we expand, because the perturbed metric breaks translation symmetry. Without loss of generality, it is convenient to choose the expansion point to be the mid-point between $x_1$ and~$x_2$:
\begin{equation}
\label{eq:midpoint}
x_0 = \frac{x_1 + x_2}{2}.
\end{equation}

Explicitly, the zeroth order term (the flat space Feynman propagator) is
\begin{align}
G_F^{(0)}\left(\Delta x\right) &= \int_{k} \frac{-i}{k^2 - i\epsilon} e^{-i k \cdot (x_1 -x_2)},\\
&= I_1\left(\Delta x\right),
\end{align}
where $I_1\left(\Delta x\right)$ is defined in Eq.~\eqref{eq:GenericI}.
The first-order correction is then given by
\begin{equation}
G_F^{(1)}\left(\Delta x, x_0\right) = -\frac{i}{2}\int_{k, k'} \frac{k_\mu k'_\nu + k'_\mu k_\nu}{\left(k^2 - i\epsilon\right)\left(k'^2 - i\epsilon\right)} {\bar{h}}^{\mu\nu}\left(k' - k\right) e^{-ikx_1} e^{ik' x_2}.
\end{equation}
Since the final result will depend on $\Delta x$ and $x_0$, it is convenient to switch to these variables from the original coordinates $x_1$ and $x_2$. As a result, 
\begin{equation}
G_F^{(1)}\left(\Delta x, x_0\right) = -\frac{i}{2}\int_{p, q} \frac{k_\mu k'_\nu + k'_\mu k_\nu}{\left(k^2 - i\epsilon\right)\left(k'^2 - i\epsilon\right)} {\bar{h}}^{\mu\nu}\left(q\right) e^{ip\Delta x} e^{iqx_0},
\end{equation}
where we defined new momenta as
\begin{equation}
k'^\mu = p^\mu + \frac{1}{2} q^\mu \qquad \text{and} \qquad k^\mu = p^\mu - \frac{1}{2} q^\mu.
\end{equation}
The Jacobian of this transformation is one. Physically, $q$ is the wavenumber carried by the external gravitational field, which we take to be small, and $p$ is a hard momentum compared to $q$. We can then transform the remaining integrand and expand to second order in $\left(q/p\right)$, yielding
\begin{equation}
\begin{aligned}
\label{eq:ExpandedStep}
G_F^{(1)}\left(\Delta x, x_0\right) = \int_{p, q} &\frac{-i}{\left(p^2 - i\epsilon\right)^2}\Bigg(p_\mu p_\nu - \frac{1}{4}q_\mu q_\nu - \frac{q^2 p_\mu p_\nu}{2\left(p^2 - i\epsilon\right)} \\
&+ \frac{\left(p \cdot q\right)^2 p_\mu p_\nu}{\left(p^2 - i\epsilon\right)^2}\Bigg) {\bar{h}}^{\mu\nu}\left(q\right) e^{ip\Delta x} e^{iqx_0} + \mathcal{O}\left(\left(q/p\right)^3\right).
\end{aligned}
\end{equation}
The $p$ and $q$ integrals are now factorized and, in this form, they can be readily solved in terms of derivatives of $I_\alpha\left(\Delta x\right)$, as defined in Eq.~\eqref{eq:GenericI}, and of $\bar{h}_{\mu\nu}\left(x_0\right)$:
\begin{equation}
\begin{aligned}
G_F^{(1)}\left(\Delta x, x_0\right) &= -\bar{h}^{\mu\nu}\left(x_0\right)\partial_\mu \partial_\nu I_2\left(\Delta x\right) + \frac{1}{4} \partial_\mu \partial_\nu \bar{h}^{\mu\nu}\left(x_0\right) I_2\left(\Delta x\right)\\
&\hspace{0.5cm} - \frac{1}{2}\Box \bar{h}^{\mu\nu}\left(x_0\right)\partial_\mu \partial_\nu I_3\left(\Delta x\right) - \partial^\rho\partial^\sigma \bar{h}^{\mu\nu}\left(x_0\right)\partial_\rho \partial_\sigma \partial_\mu \partial_\nu I_4\left(\Delta x\right).
\end{aligned}
\end{equation}
Summing the zeroth-order and first-order results, plugging in Eq.~\eqref{eq:GenericIresult} for $I_\alpha$, and rewriting combinations of $h_{\mu\nu}$ in terms of the linearized Ricci tensor and scalar where possible,
we find
\begin{equation}
\begin{aligned}
G_F(x_1,x_2) = \frac{\Gamma\left(\n/2 - 1\right)}{4\pi^{\n/2}\left(\Delta x^2 + i\epsilon\right)^{\n/2 - 2}}&\Bigg(\frac{1}{\Delta x^2 + i\epsilon}\left(1 + \frac{\Delta x^\mu \Delta x^\nu}{12}R_{\mu\nu}\left(x_0\right)\right) \\
& + \frac{1}{12\left(\n - 4\right)}R\left(x_0\right) \\
& - \frac{\left(\n - 2\right)\Delta x^\mu \Delta x^\nu}{2\left(\Delta x^2 + i\epsilon\right)^2}h_{\mu\nu}\left(x_0\right) \\
& - \frac{\left(\n - 2\right)\Delta x^\mu \Delta x^\nu \Delta x^\rho \Delta x^\sigma}{48\left(\Delta x^2 + i\epsilon\right)^2}\partial_\rho \partial_\sigma h_{\mu\nu}\left(x_0\right) \Bigg)+ \dots
\end{aligned}
\end{equation}

This result should be expressible in terms of geometrically covariant objects. For an object depending on two points, such as a two-point function, tensorial quantities defined at one point are not sufficient, and one needs to use biscalar invariants such as the squared geodesic distance $s^2\left(x_1, x_2\right)$, and the van Vleck determinant $\mathcal{D}\left(x_1, x_2\right)$ \cite{Poisson:2011nh}. These objects and their expansion in the coincident limit are reviewed in Appendix~\ref{a:GeometricInvariants}.
For our purposes here, suffice it to say that they are regular for $x_1 \to x_2$.

Using Eq.~\eqref{eq:SigmaExp} and Eq.~\eqref{eq:vvexpansion}, we find that the singular (i.e. non-analytic) part of the two-point function in the coincident limit $x_1\to x_2$ can be expressed as
\begin{equation}
\label{eq:minimalGF}
G_F(x_1,x_2) = \frac{\Gamma\left(\n/2 - 1\right)}{4\pi^{\n/2}\left(s^2\left(x_1, x_2\right) + i\epsilon\right)^{\n/2 - 2}}\left(\frac{\mathcal{D}^{1/2}\left(x_1, x_2\right)}{s^2\left(x_1, x_2\right) + i\epsilon} + \frac{1}{12\left(\n - 4\right)}R\left(x_0\right)\right) \, .
\end{equation}
Taking the $\n \to 4$ limit, we get
\begin{equation}
\lim_{\n \rightarrow 4} G_F(x_1,x_2) = \frac{1}{4\pi^2}\left(\frac{\mathcal{D}^{1/2}\left(x_1, x_2\right)}{s^2\left(x_1,x_2\right) + i\epsilon} - \frac{1}{24}R\left(x_0\right)\ln\left(s^2\left(x_1, x_2\right) + i\epsilon\right) \right),
\end{equation}
From this result we can easily read off the functions $u$ and $v$ defined in~Eq.~\eqref{eq:Hadamard}, in the coincident limit and at leading order in curvature and derivatives. This expression matches the result of Ref.~\cite[eq. (2.16)]{DeWitt:1960fc}, after taking into account the different conventions.\footnote{Our definition of the Ricci scalar differs from that of Ref.~\cite{DeWitt:1960fc} by a minus sign, and our $s^2$ is twice the $\sigma$ of Ref.~\cite{DeWitt:1960fc}. Moreover the normalization of the Green's function in Ref.~\cite{DeWitt:1960fc} is twice the one used in this work.} 

It is also interesting to look at the two-dimensional case, where minimal coupling is equivalent to conformal coupling. There we find
\begin{align}
\label{eq:2dimanswer}
\lim_{\n \rightarrow 2} G_F(x_1,x_2)&= -\frac{1}{4\pi}\left(\mathcal{D}^{1/2}\left(x_1, x_2\right) - \frac{s^2\left(x_1, x_2\right)}{24}R\left(x_0\right)\right)\ln\left(s^2\left(x_1, x_2\right) + i\epsilon\right) \nonumber \\
 &= -\frac{1}{4\pi} \ln\left(s^2\left(x_1, x_2\right) + i\epsilon\right)  \;,
\end{align}
where in the second equality we worked at leading order in curvature.
This result is consistent with the one obtained performing a Weyl transformation of the Green's function in flat space, see Appendix~\ref{a:Weyl} for more details.

In order to extend our results to the non-minimally coupled case, we add to Eq.~\eqref{eq:freeAction} a coupling to the Ricci scalar:
\begin{equation}
\label{eq:ConformalCouplingAction}
S_{\xi} = -\frac{\xi}{2}\int R\left(x\right)\phi^2\left(x\right) \sqrt{-g}\dd[\n]{x}.
\end{equation}
Since we expand the action to first order in $h$ and second derivatives of $h$, and the Ricci scalar's leading term appears at this order, $R(x)$ can be replaced by its value at the expansion point~$x_0$. Hence,
\begin{equation}
S_{\xi} = -\frac{\xi}{2}R\left(x_0\right)\int \phi^2\left(x\right)\dd[\n]{x}.
\end{equation}
Therefore, the correction to the two-point function is 
\begin{align}
\expval{T\phi\left(x_1\right)\phi\left(x_2\right)}_\xi &= -i\xi R\left(x_0\right) \int_k \frac{\left(-i\right)^2}{\left(k^2 - i\epsilon\right)^2} e^{ik\Delta x},\\
&= -\xi R\left(x_0\right) I_2\left(\Delta x\right),\\
&= \frac{\Gamma\left(\n/2 - 1\right)}{4\pi^{\n/2}\left(\Delta x^2 + i\epsilon\right)^{\n/2 - 2}}\left(-\frac{\xi}{2\left(\n - 4\right)} R\left(x_0\right)\right),
\end{align}
once again using Eq.~\eqref{eq:GenericIresult}. Adding this to the result for the minimally coupled scalar, we find
\begin{equation}\label{eq:nonminimalGF}
G_F(x_1,x_2) = \frac{\Gamma\left(\n/2 - 1\right)}{4\pi^{\n/2}\left(s^2\left(x_1, x_2\right) + i\epsilon\right)^{\n/2 - 2}}\left(\frac{\mathcal{D}^{1/2}\left(x_1, x_2\right)}{s^2\left(x_1, x_2\right) + i\epsilon} + \frac{1 - 6\xi}{12\left(\n - 4\right)}R\left(x_0\right)\right)\, .
\end{equation}

Note that, in four dimensions, conformal coupling corresponds to $\xi = 1/6$. In this case, for $d \rightarrow 4$ there is \textit{no} logarithmic singularity (no Hadamard tail), differently from the case of the minimally coupled scalar. This can be interpreted as a consequence of conformal invariance, or, to be more precise, Weyl invariance.\footnote{Conformal invariance implies Weyl invariance in unitary CFTs in $\n\leq10$, up to contact terms~\cite{Farnsworth:2017tbz}. The Weyl anomaly in $d=4$ involves quadratic curvature invariants and cannot affect this argument.} If present, the Hadamard tail would be present also for conformally flat spaces with $R\neq 0$. However, correlators in conformally flat backgrounds can be computed by a Weyl rescaling of the flat space result and do not have such a term. We provide a detailed derivation of this in Appendix~\ref{a:Weyl}. It is interesting to note that such a cancellation does not happen in any other number of dimensions, even if conformal coupling is chosen. Indeed, conformal coupling in $d$ dimensions corresponds to 
\begin{equation}\label{eq:conformalxi}
\xi_{\rm conf} = \dfrac{d-2}{4(d-1)} \; ,
\end{equation}
and the non-analytic term with the Ricci scalar does not vanish unless $\xi = 1/6$.\footnote{In $d=4$, with conformal coupling, the terms proportional to the Ricci scalar do not vanish, but are analytic for $x_1 \to x_2$.} For $d=3$ or $d>4$, however, this gravitational correction no longer corresponds to a logarithmic tail. When $d = 2$, minimal coupling and conformal coupling coincide and we recover~Eq.~\eqref{eq:2dimanswer}.

As a final note, in this section we kept all derivatives of $h_{\mu\nu}\left(x_0\right)$ up to second order. However, we can always work in locally inertial coordinates where $h_{\mu\nu}\left(x_0\right) = \partial_\rho h_{\mu\nu}\left(x_0\right) = 0$. Therefore, the first non-trivial order in derivatives of $h_{\mu\nu}$ is second-order, which corresponds to terms that go as $q^2$ in the integral. Therefore, to simplify future calculations, we will isolate terms of order $\left(q/p\right)^2$ instead of expanding up to second order in $q/p$. 

\section{Maxwell two-point function: massless spin-1}
In the previous section, we outlined an approach to computing the \textit{universal} non-analytic gravitational corrections to the flat space Green's function. To summarize, we coupled the field, $\phi$, to gravity through the stress-energy tensor, computed the correction to the two-point function following Eq.~\eqref{eq:general}, and then enforced general covariance. Our results then matched the known Hadamard tail. Our diagrammatic approach allows us to compute universal gravitational corrections for more general situations. In this section, we will compute the two-point function for a massless spin-1 field, $A_\mu\left(x\right)$, in a curved background. 

To begin, in a general $R_\xi$ gauge, the free Maxwell action is given by
\begin{equation} \label{eq:spin1action}
S = \int \left(-\frac{1}{4} F_{\mu \nu} F^{\mu\nu} - \frac{1}{2\alpha} \left(\nabla_\mu A^\mu\right)^2\right) \sqrt{-g} \dd[d]{x},
\end{equation}
where we denote the gauge parameter by $\alpha$ to avoid conflict with the non-minimal coupling to the Ricci scalar previously defined. The flat space stress-energy tensor is
\begin{equation}
\begin{aligned}
T_{\mu\nu} &= F_{\mu\rho} F_\nu^{\hphantom{\nu}\rho} - \frac{1}{4} \eta_{\mu\nu} F_{\rho\sigma} F^{\rho\sigma} \\
&\hspace{0.4cm} - \frac{1}{\alpha}\left[A_\mu \partial_\nu\partial_\rho A^\rho + A_\nu \partial_\mu \partial_\rho A^\rho - \eta_{\mu\nu} A^\rho \partial_\rho \partial_\sigma A^\sigma - \frac{1}{2} \eta_{\mu\nu} \partial_\rho A^\rho \partial_\sigma A^\sigma\right].
\end{aligned}
\end{equation}
The first term is the Maxwell stress-energy tensor while the second line is the contribution from the gauge fixing piece.

The Feynman rule describing the propagator is given by 
\begin{equation}
\label{eq:FlatSpace Spin 1 Prop}
\begin{gathered}
\begin{tikzpicture}
\draw[line width=0.3mm, decorate, decoration={snake}] (-0.75, 0) -- (0.75, 0); 
\end{tikzpicture}
\end{gathered} \: \: = \frac{-i}{k^2 - i\varepsilon}\left(\eta_{\mu\nu} - \left(1 - \alpha\right)\frac{k_\mu k_\nu}{k^2}\right).
\end{equation}
Meanwhile, the leading interaction vertex is given by 
\begin{equation}
\begin{aligned}
\begin{gathered}
\begin{tikzpicture}
\draw[line width=0.3mm, decorate, decoration={snake}] (-1.25, 0) -- (-0.1, 0);
\draw[line width=0.3mm, decorate, decoration={snake}] (0.1, 0) -- (1.25, 0);
\draw[line width=0.3mm] (0, 0) node[cross=3.75pt] {};
\draw[line width=0.3mm] (0, 0) circle (3.75pt);
\end{tikzpicture}
\end{gathered} \: \: &= \frac{i}{2}h^{\rho\sigma}\bigg(\left(\eta_{\mu\rho} \eta_{\nu\sigma} + \eta_{\mu \sigma} \eta_{\nu \rho} - \eta_{\mu\nu} \eta_{\rho\sigma}\right) k \cdot k^\prime + \eta_{\rho\sigma} k_\nu k_\mu^\prime - \eta_{\sigma\nu} k_\rho k^\prime_\mu \\
&\hspace{1.5cm}  - \eta_{\rho\nu} k_\sigma k_\mu^\prime -\eta_{\sigma \mu} k_\nu k_\rho^\prime + \eta_{\mu\nu} k_\sigma  k_\rho^\prime - \eta_{\rho\mu} k_\nu k^\prime_\sigma + \eta_{\mu\nu} k_\rho k_\sigma^\prime \\
&\hspace{1.5cm} + \frac{1}{\alpha}\big(-\eta_{\rho\sigma} k_\mu k_\nu + \eta_{\sigma \nu} k_\mu k_\rho + \eta_{\rho\nu} k_\mu k_\sigma + \eta_{\rho\sigma} k_\mu k_\nu^\prime \\
&\hspace{2.5cm} - \eta_{\rho\sigma} k_\mu^\prime k_\nu^\prime + \eta_{\mu\sigma} k_\nu^\prime k_\rho^\prime + \eta_{\rho \mu} k_\nu^\prime k_\sigma^\prime \big) \bigg)  \, .
\end{aligned}
\end{equation}
With these, we can compute the two-point function for the vector
\begin{equation}
    G_{\mu\nu}(x_1,x_2)\equiv \expval{TA_\mu\left(x_1\right)A_\nu\left(x_2\right)}_{g} = \:\: \begin{gathered}
\begin{tikzpicture}
\draw[line width=0.3mm, decorate, decoration={snake}] (-0.75, 0) -- (0.75, 0); 
\filldraw (-0.75, 0) circle (0.075cm);
\node at (-0.75, 0.25) {$x_1$};
\filldraw (0.75, 0) circle (0.075cm);
\node at (0.75, 0.25) {$x_2$};
\end{tikzpicture}
\end{gathered} \: \: + \: \: \begin{gathered}
\begin{tikzpicture}
\draw[line width=0.3mm, decorate, decoration={snake}] (-1.25, 0) -- (-0.1, 0);
\draw[line width=0.3mm, decorate, decoration={snake}] (0.1, 0) -- (1.25, 0);
\draw[line width=0.3mm] (0, 0) node[cross=3.75pt] {};
\draw[line width=0.3mm] (0, 0) circle (3.75pt);
\filldraw (-1.25, 0) circle (0.075cm);
\filldraw (1.25, 0) circle (0.075cm);
\node at (-1.25, 0.25) {$x_1$};
\node at (1.25, 0.25) {$x_2$};
\end{tikzpicture}
\end{gathered} \: \: + \dots.
\end{equation}
As before, the momentum space integrals can be simplified by expanding them to order $\left(q/p\right)^2$ in the short distance approximation. Then, organizing the result in manifestly covariant form using the identities in Appendix~\ref{a:GeometricInvariants}, gives
\begin{equation}
\label{eq:Spin1Split}
G_{\mu\nu}\left(x_1, x_2\right) = {\cal G}_{\mu\nu}\left(x_1, x_2\right) + \left(1 - \alpha\right)\nabla_\mu^{(1)} \nabla_\nu^{(2)} \Phi\left(x_1, x_2\right),
\end{equation}
where
\begin{subequations}
\begin{align}
{\cal G}_{\mu\nu}\left(x_1, x_2\right) &= \frac{\Gamma\left(d/2 - 1\right)}{4\pi^{d/2}\left(s^2 + i\varepsilon\right)^{d/2 - 1}}\left[g_{\mu\nu} \mathcal{D}_{12}^{1/2} - \frac{s^2}{2\left(d - 4\right)}\left(R_{\mu\nu} - \frac{1}{6} R \, g_{\mu\nu}\right)\right]\\
\Phi\left(x_1, x_2\right) &= -\frac{\Gamma\left(d/2 - 1\right)}{8\pi^{d/2}\left(d - 4\right)\left(s^2 + i\varepsilon\right)^{d/2 -2}}\left[\mathcal{D}^{1/2}_{12} + \frac{s^2}{12\left(d - 6\right)} R\right] \; .
\end{align}
\end{subequations}
with  $\mathcal{D}_{12} \equiv \mathcal{D}\left(x_1, x_2\right)$.
Note that the $\Phi$ piece drops out in the Feynman gauge, $\alpha = 1$. In that gauge and in $d = 4$, we recover the result found in Ref.~\cite{DeWitt:1960fc}. The split in Eq.~\eqref{eq:Spin1Split} is not an artifact of the short-distance expansion: it was shown in Ref.~\cite{Frob:2017gez} that it holds in any number of dimensions and in a general covariant gauge. Just as they point out, our $\Phi$ is proportional to the scalar Hadamard distribution. We also check that this result matches theirs in $d = 4$ as well as Ref.~\cite{Toms:2014tia} in general dimensions at linear order in curvature.

More generally, the $\Phi$ piece in Eq.~\eqref{eq:Spin1Split} is the curved space counterpart of the longitudinal term $k_\mu k_\nu$ in the flat space propagator in Eq.~\eqref{eq:FlatSpace Spin 1 Prop}. Hence, we expect it to drop out of any gauge invariant observable. Explicitly, for the field strength correlator the gauge dependence drops out, since covariant derivatives acting on $\Phi$ commute,  leaving
\begin{equation} \label{eq:spin1FF}
\expval{F_{\mu\nu}\left(x_1\right) F_{\rho\sigma}\left(x_2\right)} = 4\nabla^{(1)}_{[\mu} \nabla^{(2)}_{[\rho} G_{\nu]\sigma]}^{(0)}\left(x_1, x_2\right).
\end{equation}
Carrying out the derivatives, we obtain
\begin{equation}
\begin{aligned}
\expval{F_{\mu\nu}\left(x_1\right) F_{\rho\sigma}\left(x_2\right)} &= \frac{\Gamma\left(d/2 - 1\right)}{\pi^{d/2}} \partial^{(1)}_{[\mu} \partial^{(2)}_{[\rho} \left[\frac{g_{\nu]\sigma]} \mathcal{D}_{12}^{1/2}}{\left(s^2 + i\varepsilon\right)^{d/2 - 1}}\right] \\
&\quad - \frac{\Gamma\left(d/2 - 1\right)}{8\pi^{d/2}\left(s^2 + i\varepsilon\right)^{d/2 - 1}} \bigg[\left(g_{\mu \rho} - \left(d - 2\right)\frac{\sigma_\mu \sigma_\rho}{s^2}\right)\left(R_{\nu\sigma} - \frac{1}{6} R g_{\nu\sigma}\right) \\
&\hspace{4cm} - \left(\rho \leftrightarrow \sigma\right) - \left(\mu \leftrightarrow \nu\right) + \left(\mu, \rho \leftrightarrow \nu, \sigma\right)\bigg],
\end{aligned}
\end{equation}
where $\sigma_\mu = \nabla_\mu (s^2/2)$. As opposed to $G_{\mu\nu}$, this has no logarithmic (Hadamard) tail in $d = 4$. This is similar to what occurred in the case of the conformally coupled free scalar in Section~\ref{s:spin0}. Notice however that, differently from the conformally coupled scalar, power-law curvature corrections survive in the $d\to 4$ limit. As we will see later, the same happens also for scalar primaries with scaling dimensions above the unitarity bound.

\section{Graviton two-point function: spin-2 in a curved background}\label{spin 2}

As a final example, we turn our attention to a massless spin-2 field, $\delta g_{\mu\nu}$, on a curved background. We take the action to be the pure Einstein-Hilbert term and expand the metric as $g_{\mu\nu} = \bar{g}_{\mu\nu} + \delta g_{\mu\nu}$. To simplify the notation we will drop the bar from the background values, and use $\delta g_{\mu\nu}$ to refer to the perturbations. Including the de Donder gauge fixing term
\begin{equation}
S_{\text{GF}} = \int \left(\nabla^\mu \delta g_{\mu\nu} - \frac{1}{2} \nabla_\nu \delta g\right)^2 \sqrt{-g} \dd[4]{x},
\end{equation}
the action is \cite{Allen:1987bn}
\begin{equation}
\label{eq:spin 2 action}
\begin{aligned}
S &= \int \bigg[\frac{1}{2} \delta g^{\mu\nu} \Box \delta g_{\mu\nu} - \frac{1}{4} \delta g \Box \delta g + R_{\mu\nu\rho\sigma} \delta g^{\mu\nu} \delta g^{\rho\sigma} + R^{\nu\rho} \delta g^\mu_{\hphantom{\mu}\nu} \delta g_{\mu\rho} \\
&\hspace{1.25cm} - R_{\mu\nu} \delta g^{\mu\nu} \delta g - \frac{1}{2} R \delta g_{\mu\nu} \delta g^{\mu\nu} + \frac{1}{4} R \delta g^2
\bigg] \sqrt{-g} \dd[4]{x}. 
\end{aligned}
\end{equation}
There are also contributions to the action from ghosts. In our chosen gauge, ghosts couple quadratically to the graviton and hence only contribute at loop order. Differently from the spin-1 case, the gauge fixing parameter is fixed rather than kept arbitrary, to simplify the expressions.
The flat space stress-energy tensor is 
\begin{equation}
\begin{aligned}
T_{\mu\nu}
={}& \delta g_{\mu}{}^{\rho}\,\partial_{\rho}\partial_{\sigma} \delta g_{\nu}{}^{\sigma}
  + \delta g_{\nu}{}^{\rho}\,\partial_{\rho}\partial_{\sigma} \delta g_{\mu}{}^{\sigma}
  + \delta g_{\mu}{}^{\rho}\,\partial_{\sigma}\partial_{\rho} \delta g_{\nu}{}^{\sigma}
  + \delta g_{\nu}{}^{\rho}\,\partial_{\sigma}\partial_{\rho} \delta g_{\mu}{}^{\sigma}\\
 &- \delta g_{\mu}{}^{\rho}\,\partial_{\rho}\partial_{\nu} \delta g
  - \delta g_{\nu}{}^{\rho}\,\partial_{\rho}\partial_{\mu} \delta g
  - 2\,\delta g_{\mu}{}^{\rho}\,\Box \delta g_{\nu\rho}
  - 2\,\delta g_{\nu}{}^{\rho}\,\Box \delta g_{\mu\rho}\\[4pt]
 &+ 2\,\partial_{\rho} \delta g \,\partial^{\rho}\delta g_{\mu\nu}
  - 4\,\partial^{\rho} \delta g_{\mu\nu}\,\partial^{\sigma} \delta g_{\rho\sigma}
  - 2\,\delta g^{\rho\sigma}\,\partial_{\sigma}\partial_{\rho} \delta g_{\mu\nu}
  + \delta g\,\Box \delta g_{\mu\nu}\\
 &- 2\,\delta g_{\mu\nu}\,\partial_{\sigma}\partial_{\rho}\delta g^{\rho\sigma}
  + 2\,\delta g_{\mu\nu}\,\Box \delta g + 2\,\partial_{\rho}\delta g_{\mu}{}^{\rho}\,\partial_{\sigma} \delta g_{\nu}{}^{\sigma}
  + 2\,\partial_{\rho}\delta g_{\nu\sigma}\,\partial^{\sigma}\delta g_{\mu}{}^{\rho} \\[4pt]
 &- 2\,\partial_{\sigma} \delta g_{\nu\rho}\,\partial^{\sigma} \delta g_{\mu}{}^{\rho} + 2\,\delta g^{\rho\sigma}\,\partial_{\sigma}\partial_{\mu} \delta g_{\nu\rho}
  + 2\,\delta g^{\rho\sigma}\,\partial_{\sigma}\partial_{\nu}\delta g_{\mu\rho}
  - \delta g\,\partial_{\sigma}\partial_{\mu}\delta g_{\nu}{}^{\sigma} \\[4pt]
 &- \delta g\,\partial_{\sigma}\partial_{\nu}\delta g_{\mu}{}^{\sigma} - \partial_{\rho}\delta g_{\nu}{}^{\rho}\,\partial_{\mu}\delta g
  - \partial_{\rho}\delta g_{\mu}{}^{\rho}\,\partial_{\nu}\delta g
  - \partial_{\rho}\delta g\,\partial_{\mu}\delta g_{\nu}{}^{\rho} \\
 & - \partial_{\rho}\delta g\,\partial_{\nu}\delta g_{\mu}{}^{\rho} + 2\,\partial_{\sigma}\delta g_{\rho}{}^{\sigma}\,\partial_{\mu}\delta g_{\nu}{}^{\rho}
  + 2\,\partial_{\sigma}\delta g_{\rho}{}^{\sigma}\,\partial_{\nu}\delta g_{\mu}{}^{\rho}
  - \delta g^{\rho\sigma}\,\partial_{\mu}\partial_{\nu}\delta g_{\rho\sigma} \\
 &- \delta g^{\rho\sigma}\,\partial_{\nu}\partial_{\mu}\delta g_{\rho\sigma} - \partial_{\mu}\delta g^{\rho\sigma}\,\partial_{\nu}\delta g_{\rho\sigma}
  + \tfrac{1}{2}\,\partial_{\mu}\delta g\,\partial_{\nu}\delta g
  + \delta g\,\partial_{\nu}\partial_{\mu}\delta g\\[4pt]
 &+ \eta_{\mu\nu}\Big[\,
    \delta g^{\rho\sigma}\,\partial_{\sigma}\partial_{\rho}\delta g
  - \delta g^{\rho\sigma}\,\partial_{\sigma}\partial^{\lambda}\delta g_{\rho\lambda}
  - \delta g^{\rho\sigma}\,\partial^{\lambda}\partial_{\sigma}\delta g_{\rho\lambda}
  + \delta g\,\partial^{\lambda}\partial^{\sigma}\delta g_{\sigma\lambda}\\
 &\hphantom{+ g_{\mu\nu}\Big[\,}
  + 2\,\delta g^{\rho\sigma}\,\Box \delta g_{\rho\sigma}
  - \delta g\,\Box \delta g
  - \partial_{\rho}\delta g^{\rho\sigma}\,\partial^{\lambda}\delta g_{\sigma\lambda}
  + 2\,\partial^{\sigma}\delta g\,\partial^{\lambda}\delta g_{\sigma\lambda}\\
 &\hphantom{+ g_{\mu\nu}\Big[\,}
  - \tfrac{3}{4}\,\partial_{\sigma}\delta g\,\partial^{\sigma}\delta g
  - \partial_{\sigma}\delta g_{\rho\lambda}\,\partial^{\lambda}\delta g^{\rho\sigma}
  + \tfrac{3}{2}\,\partial_{\lambda}\delta g_{\rho\sigma}\,\partial^{\lambda}\delta g^{\rho\sigma}
  \,\Big]
\end{aligned}
\end{equation}
Additionally, the flat space propagator is
\begin{equation}
\label{eq:spin2prop}
\begin{gathered}
\begin{tikzpicture}
\draw[line width=0.3mm, decorate, decoration={coil, aspect = 1.1, segment length= 2mm}] (-0.75, 0) -- (0.75, 0); 
\end{tikzpicture}
\end{gathered} \: \: = \frac{-i}{2\left(k^2 - i\varepsilon\right)}\left(\eta_{\mu\rho} \eta_{\nu\sigma} + \eta_{\mu\sigma} \eta_{\nu\rho} - \frac{2}{d-2}\eta_{\mu\nu}\eta_{\rho\sigma}\right).
\end{equation}
Before continuing, there are, at this point, three important distinctions compared to the earlier spin-0 and spin-1 cases: 
\begin{enumerate}[label=\textbf{\arabic*.}]
\item \textbf{The background is no longer arbitrary.} For the spin-0 and spin-1 fields, the background metric could be left completely general: the two point functions were computed without any assumption on $g_{\mu\nu}$ beyond it being weakly curved near $x_0$. For a spin-2 fluctuation, there may be additional contributions depending on whether the background satisfies the Einstein equations or not. This is because the fluctuation is part of the metric itself, $g_{\mu\nu} = \bar{g}_{\mu\nu} + \delta g_{\mu\nu}$, so linear terms in $\delta g_{\mu\nu}$ cancel in the action only if the background Einstein equations are satisfied. If they do not vanish, they act as a source for $\delta g$: the one-point function $\expval{\delta g_{\mu\nu}}$ would be nonzero, and the two-point function would receive corrections from diagrams in which a propagator ends on this linear vertex (tadpoles). Such insertions are of the same order in the curvature expansion as the corrections we consider, so they cannot be dropped. One can, however, shift the expansion point, $\delta g_{\mu\nu} \to \expval{\delta g_{\mu\nu}} + \delta g_{\mu\nu}$, until the linear term is absent -- that is, until the background solves the Einstein equations. We therefore impose them from the outset. 
\item \textbf{Independence of Background Quantities.} As a consequence of imposing the background Einstein equations, the curvature invariants appearing in the answer are no longer independent: $R_{\mu\nu}$  is fixed in terms of the background stress-energy tensor. For example, for a vacuum background with a cosmological constant, $R_{\mu\nu}= \frac{2\Lambda}{d-2} g_{\mu\nu}$, and every Ricci structure in the eventual two-point function collapses to a multiple of $\Lambda$ and combinations of Riemann tensors. We mention this as it will be relevant for comparing with the results of Ref.~\cite{Allen:1987bn}.
\item \textbf{Coupling to Other Matter.} By the equivalence principle, $\delta g$ couples to every other field. If the background is sourced by matter, say a scalar field profile $\bar{\phi}(x)$, then the vertex $\delta g \,\delta g^{\mu\nu}\partial_\mu \bar{\phi}\, \partial_\nu \bar{\phi}$ produces another contribution to the two-point function
\begin{equation}
\label{eq:Diagram Expansion spin 2}
G_{\mu\nu\rho\sigma} \equiv \langle T \delta g_{\mu\nu}\left(x_1\right) \delta g_{\rho \sigma}\left(x_2\right)\rangle_g = \: \: 
\begin{gathered}
\begin{tikzpicture}
\draw[line width=0.3mm, decorate, decoration={coil, aspect = 1.1, segment length= 2mm}] (-0.75, 0) -- (0.75, 0); 
\end{tikzpicture}
\end{gathered} \: \: + \: \: \begin{gathered}
\begin{tikzpicture}
\draw[line width=0.3mm, decorate, decoration={coil, aspect = 1.1, segment length= 2mm}] (-1.25, 0) -- (-0.1, 0);
\draw[line width=0.3mm, decorate, decoration={coil, aspect = 1.1, segment length= 2mm}] (0.1, 0) -- (1.25, 0);
\draw[line width=0.3mm] (0, 0) node[cross=3.75pt] {};
\draw[line width=0.3mm] (0, 0) circle (3.75pt);
\end{tikzpicture}
\end{gathered} \: \: + \: \: 
\begin{gathered}
\begin{tikzpicture}
\draw[line width=0.3mm, decorate, decoration={coil, aspect = 1.1, segment length= 2mm}] (-1.25, 0) -- (-0.1, 0);
\draw[line width=0.3mm, decorate, decoration={coil, aspect = 1.1, segment length= 2mm}] (0.1, 0) -- (1.25, 0);
\filldraw[line width=0.3mm] (0, 0) circle (3.75pt);
\end{tikzpicture}
\end{gathered} \: \: .
\end{equation}
Such a contribution would not be suppressed: the background equations give $\partial_\mu\bar{\phi}\partial_\nu\bar{\phi} \sim R_{\mu\nu}$, which enters at the same order in the short distance expansion as the second diagram. Unlike the first two diagrams, however, it depends on the dynamics of the sector that sources the curvature and not only on its stress-energy tensor. Moreover, any matter sector background solution admits perturbations, which can mix with gravitational perturbations. Familiar physical phenomena associated with such a mixing are scalar cosmological perturbations and Jeans instability. Keeping track of all this would require a model-dependent analysis. We thus restrict our analysis to pure-gravity cases, that is, cases with no matter.
\end{enumerate}

We will start by computing the first two terms of Eq.~\eqref{eq:Diagram Expansion spin 2}. As before, we pass to hard and soft momenta $p$ and $q$, expand to second order in $q/p$ and use the master integrals in Appendix~\ref{a:ImportantIdentities}. After re-expressing the answer in terms of covariant objects as shown in Appendix~\ref{a:GeometricInvariants}, the two point function from the first two diagrams becomes
\begin{equation}
\label{eq:Universal Spin 2}
G_{\mu\nu\rho\sigma}^{(R)} = \frac{\Gamma\left(d/2 - 1\right)}{4\pi^{d/2}\left(s^2 + i\varepsilon\right)^{d/2 - 1}}\left[\mathcal{D}^{1/2}_{12} \mathcal{P}_{\mu\nu\rho\sigma} + \frac{s^2}{d - 4} v_{\mu\nu\rho\sigma}\right],
\end{equation}
where the superscript indicates that this is not yet the full two-point function. Here, 
\begin{equation}
\label{eq:spin-2 propagator num}
\mathcal{P}_{\mu\nu\rho\sigma} = \frac{1}{2} g_{\mu\rho} g_{\nu\sigma} + \frac{1}{2} g_{\mu\sigma} g_{\nu\rho} - \frac{1}{d - 2} g_{\mu\nu} g_{\rho\sigma},
\end{equation}
is a traceless projector, as in Eq.~\eqref{eq:spin2prop}, and
\begin{equation} \label{eq:spin2vpart}
\begin{aligned}
v_{\mu\nu\rho\sigma} &= \frac{1}{2}\left(R_{\mu\rho\nu\sigma} + R_{\mu\sigma\nu\rho}\right) - \frac{1}{d - 2}\left(g_{\mu\nu} R_{\rho\sigma} + g_{\rho\sigma} R_{\mu\nu}\right) \\ 
&\hspace{0.4cm} + \frac{1}{4}\left(g_{\mu\rho} R_{\nu\sigma} + g_{\mu\sigma} R_{\nu\rho} + g_{\nu\rho} R_{\mu\sigma} + g_{\nu\sigma} R_{\mu\rho}\right) \\
&\hspace{0.4cm} - \frac{5}{24}R\left(g_{\mu\rho} g_{\nu\sigma} + g_{\mu\sigma} g_{\nu\rho} \right) + \frac{5}{12\left(d - 2\right)} R\, g_{\mu\nu} g_{\rho\sigma}. 
\end{aligned}
\end{equation}
As before, all curvature quantities are evaluated at the expansion point $x_0$. We note that this answer agrees with the one contained in Ref.~\cite{Allen:1987bn} in four dimensions, taking into account that they assume the vacuum Einstein equations with a cosmological constant throughout.

The details of the remaining piece depends on the detail of the background source -- our computation to this point captures the universal correction due to the coupling to curvature. For simplicity and as anticipated, we will consider a restricted stress-energy tensor sourced by the cosmological constant, hence the background Einstein equations are
\begin{equation}
\label{eq:bkg equations}
R_{\mu\nu} - \frac{1}{2} R g_{\mu\nu} + \Lambda g_{\mu\nu} = 0,
\end{equation}
The additional terms that appear in the action for the spin-2 perturbations, $\delta g_{\mu\nu}$ are
\begin{equation}
S_{\Lambda} = \int \Lambda \left(\delta g_{\mu\nu} \delta g^{\mu\nu} - \frac{1}{2} \delta g^2\right) \sqrt{-g} \dd[d]{x}. 
\end{equation}
From the background equations of motion $\Lambda = \left(d - 2\right)R/2$, hence $\Lambda$ is already of first order in the curvature. In fact the action $S_\Lambda$ is the same, up to a proportionality factor, as the last two terms in Eq.~\eqref{eq:spin 2 action}, i.e. the terms in the action with an overall Ricci scalar $R$. Hence, the computation is the same as that for those two terms and gives an additional contribution to the two-point function
\begin{equation}
G_{\mu\nu\rho\sigma}^{\left(\Lambda\right)} = \frac{\Gamma\left(d/2 - 2\right)\Lambda \mathcal{P}_{\mu\nu\rho\sigma}}{8\pi^{d/2}\left(s^2 + i\varepsilon\right)^{d/2 - 2}},
\end{equation}
where $\mathcal{P}_{\mu\nu\rho\sigma}$ is the sum of spin-2 polarizations given in Eq.~\eqref{eq:spin-2 propagator num}.

The total two-point function is then given by 
\begin{align}\label{eq:spin2G}
G_{\mu\nu\rho\sigma} & = G^{\left(R\right)}_{\mu\nu\rho\sigma} + G^{\left(\Lambda\right)}_{\mu\nu\rho\sigma} \\
& = \frac{\Gamma\left(d/2 - 1\right)}{4\pi^{d/2}\left(s^2 + i\varepsilon\right)^{d/2 - 1}}\left[\left(\mathcal{D}^{1/2}_{12} + \frac{s^2 \Lambda}{d - 4}\right)\mathcal{P}_{\mu\nu\rho\sigma} + \frac{s^2}{d - 4} v_{\mu\nu\rho\sigma}\right] \; .
\end{align}
Using the background equations of motion, Eq.~\eqref{eq:bkg equations}, the second term greatly simplifies:
\begin{equation}\label{eq:spin2V}
\begin{aligned}
v_{\mu\nu\rho\sigma} &= \frac{1}{2}\left(R_{\mu\rho\nu\sigma} + R_{\mu\sigma\nu\rho}\right) -\frac{\left(5d - 12\right)\Lambda}{12\left(d - 2\right)}\left(g_{\mu\rho} g_{\nu\sigma} + g_{\mu\sigma} g_{\nu\rho}\right) + \frac{\left(5d - 24\right)\Lambda}{6\left(d - 2\right)^2}g_{\mu\nu} g_{\rho\sigma}. 
\end{aligned}
\end{equation}
In the $d \rightarrow 4$ limit, the terms with a pole in $d - 4$ contribute to the coefficient of the logarithm in the resulting Hadamard distribution:
\begin{equation}
\lim_{d \rightarrow 4} G_{\mu\nu\rho\sigma} = \frac{1}{4\pi^2}\left[\frac{\mathcal{D}_{12}^{1/2}}{s^2 + i\varepsilon} - \frac{1}{2}\left(\Lambda \mathcal{P}_{\mu\nu\rho\sigma} + v_{\mu\nu\rho\sigma} \right)\log\left(s^2 + i\varepsilon\right)\right].
\end{equation}
As before, we only keep the terms that are non-analytic in the coincident limit.

\section{Scalar two-point functions: primaries in CFTs}
\label{s:Conformal Primaries}

In the previous sections, we discussed how to compute gravitational corrections to OPE coefficients of free or weakly coupled fields. 
An interesting generalization of this result is to generic two-point functions of conformal primary fields in strongly coupled CFTs, where conformal invariance completely constrains the form of two- and three-point functions. For a review and possible physical applications, see Ref.~\cite{Poland:2018epd}. In this section, focusing mainly on CFTs in $d > 2$, we compute the leading-order gravitational corrections to the two-point correlation function for a generic conformal primary scalar, $\mathcal{O}_\Delta\left(x\right)$, of arbitrary scaling dimension ${\Delta \geq (d-2)/2}$.  In this case we work directly in position space. Moreover, for simplicity, we will work in Euclidean space, so we will drop the time-ordered products and the $i\varepsilon$.\footnote{For simplicity of notation we will keep denoting the flat space metric with $\eta_{\mu\nu}$, even if in this case it is a Euclidean metric $(+,\dots,+)$.} 

We stress that we are coupling a flat space local CFT to a background gravitational field, a standard procedure in the literature, see for instance Ref.~\cite{Osborn:1993cr}. This coupling generically breaks conformality, through the Weyl anomaly. Our formalism applies equally well and the only reason we consider flat space CFTs, rather than more general QFTs, is because we can take advantage of closed form analytical expressions. In Appendix~\ref{a:anomaly} we show how our formalism can be used to provide an alternative derivation of the Weyl anomaly in 2D local CFTs with generic central charge.

Recall from Eq.~\eqref{eq:general} that the correlation function of two operators in a curved background can be expanded near the coincident limit, $x_1 \rightarrow x_2$, and expressed as a series in gravitational corrections away from flat-spacetime,
\begin{equation}
\expval{\mathcal{O}_{\Delta}\left(x_1\right) \mathcal{O}_{\Delta'}\left(x_2\right)}_g = \expval{\mathcal{O}_{\Delta}\left(x_1\right) \mathcal{O}_{\Delta'}\left(x_2\right)}_0 + \expval{\mathcal{O}_{\Delta}\left(x_1\right) \mathcal{O}_{\Delta'}\left(x_2\right)}_1 + \dots,
\end{equation}
where the subscripts indicate the order in metric perturbation $h$. The lowest order term is the flat space two-point function, which is fixed by conformal invariance to be
\begin{equation}
\expval{\mathcal{O}_{\Delta}\left(x_1\right) \mathcal{O}_{\Delta'}\left(x_2\right)}_0 = \frac{C_d}{\left(\Delta x\right)^{2\Delta}} \, \delta_{\Delta \Delta'} \, . 
\end{equation}
The normalization of the two-point function is kept arbitrary: $C_d \equiv 1$ corresponds to the one usually adopted in the CFT literature; $C_d \equiv \Gamma\left(d/2 - 1\right) / (4\pi^{d/2})$ corresponds to the canonical normalization of a free scalar field theory, and matches the convention we used in the diagrammatic approach. 

As before, the first-order term is itself expressible as the integral of a flat-space three-point correlation function with the background metric:
\begin{equation}
\label{eq:GenericCorrection}
\expval{\mathcal{O}_{\Delta}\left(x_1\right) \mathcal{O}_{\Delta'}\left(x_2\right)}_1 = \frac{1}{2}\int h_{\mu\nu}\left(y\right) \expval{\mathcal{O}_{\Delta}\left(x_1\right) T^{\mu\nu}\left(y\right) \mathcal{O}_{\Delta'}\left(x_2\right)}_0 \dd[d]{y}.
\end{equation}
The three-point correlator in flat space is also fixed by conformal invariance, up to contact terms. Using the fact that the scaling dimension of the stress-energy tensor is the spacetime dimension, we have \cite{Simmons-Duffin:2016gjk}
\begin{align}
\label{eq:3ptCF}
\expval{\mathcal{O}_{\Delta}\left(x_1\right) T^{\mu\nu}\left(y\right) \mathcal{O}_{\Delta'}\left(x_2\right)}_0 = &-\frac{\Delta C_d \,\Gamma\left(d/2+1\right)}{\left(d - 1\right)\pi^{d/2} \left(\Delta x\right)^{2\Delta - 2\Delta_0}} \frac{z^\mu z^\nu - \left(\eta^{\mu\nu}/d\right) z^2}{\left(x_1 - y\right)^{2\Delta_0} \left(y - x_2\right)^{2\Delta_0}} \delta_{\Delta \Delta'} \nonumber \\ 
& + \mbox{contact terms}
\end{align}
where $\Delta_0 = d/2 - 1$ is the free scalar scaling dimension and 
\begin{equation}
\label{eq:z_def}
z^\mu = \frac{\left(x_1 - y\right)^\mu}{\left(x_1 - y\right)^2} + \frac{\left(y - x_2\right)^\mu}{\left(y - x_2\right)^2}.
\end{equation}
The contact terms depend on the definition of the stress-energy tensor and will be fixed later. 

Since we are interested in the leading singular behavior for $x_1 \to x_2$, by the equivalence principle we can also expand the metric to second derivatives around the midpoint Eq.~\eqref{eq:midpoint},
\begin{equation}
h_{\mu\nu}\left(y\right) \simeq \frac{1}{2}\left(y - x_0\right)^\rho \left(y - x_0\right)^\sigma \partial_\rho \partial_\sigma h_{\mu\nu}\left(x_0\right),
\end{equation}
where we used local inertial coordinates $h_{\mu\nu}\left(x_0\right) = \partial_\rho h_{\mu\nu}\left(x_0\right) = 0$. This is equivalent to the approximation used in the previous section in which we expanded to second order in $\left(q/p\right)$. It is convenient to define a traceless metric perturbation 
\begin{equation}
\hat{h}_{\mu\nu} \equiv h_{\mu\nu} - \frac{1}{d}\eta^{\mu\nu} h \,.
\end{equation}
In terms of this, we find that the first order correction to the two-point function, up to contact terms, is given by:
\begin{equation}
\begin{aligned}
\expval{\mathcal{O}_{\Delta}\left(x_1\right) \mathcal{O}_{\Delta'}\left(x_2\right)}_1 &= -\frac{2^d \pi^{d/2} \Delta C_d \,\Gamma\left(d/2+1\right)}{4\left(d - 1\right)\left(\Delta x\right)^{2\Delta -2 \Delta_0}} \delta_{\Delta \Delta'} \partial_\rho \partial_\sigma \hat{h}_{\mu\nu}\left(x_0\right) \\
&\hspace{0.4cm} \times \int \frac{z^\mu z^\nu\left(y - x_0\right)^\rho \left(y - x_0\right)^\sigma}{\left(x_1 - y\right)^{2\Delta_0} \left(y - x_2\right)^{2\Delta_0}} \frac{\dd[d]{y}}{\left(2\pi\right)^d} + \mbox{contact terms}
\end{aligned}
\end{equation}
The integral is tedious but straightforward to evaluate using the Feynman parameterization Eq.~\eqref{eq:SymmetricFeynmanParameterization}, the reduction of Euclidean invariant tensor integrals to scalar integrals, \cite[eq. (4.3.11)]{Collins:1984xc}, and the evaluation of those scalar integrals in arbitrary dimensions \cite[eq. (14.27)]{Srednicki:2007qs}. The result, expressed in terms of the Ricci scalar, $R$, and the Ricci tensor, $R_{\mu\nu}$, is
\begin{equation}
\begin{aligned}
\expval{\mathcal{O}_{\Delta}\left(x_1\right) \mathcal{O}_{\Delta'}\left(x_2\right)}_{1} &= -\frac{C_d\left(\Delta/\Delta_0\right)}{24 \left(d - 1\right)\left(\Delta x\right)^{2\Delta - 2}}\delta_{\Delta \Delta'} \times \\
&\hspace{0.4cm} \Bigg(R\left(x_0\right) - \frac{2\left(d - 1\right)}{(\Delta x)^2} \Delta x^\mu \Delta x^\nu\left(R_{\mu\nu}\left(x_0\right) + \frac{3\Delta_0}{2d}\nabla_\mu \nabla_\nu h\left(x_0\right)\right) \\
&\hspace{0.6cm} + \frac{\left(d - 1\right)\Delta_0}{\left(\Delta x\right)^4} \Delta x^\mu \Delta x^\nu \Delta x^\rho \Delta x^\sigma \partial_\rho \partial_\sigma h_{\mu\nu}\left(x_0\right)\Bigg) + \mbox{contact terms}
\end{aligned}
\end{equation}

Let us now determine the contact terms. We can fix those by requiring that the stress-energy tensor is such that its trace generates scaling transformation, which at the level of correlation functions implies the Ward identity
\begin{equation}
\expval{\mathcal{O}_{\Delta}\left(x_1\right) T^\mu_{\hphantom{\mu}\mu}\left(y\right) \mathcal{O}_{\Delta'}\left(x_2\right)}_0 = \Delta\left(\delta^{(d)}\left(y - x_1\right) + \delta^{(d)}\left(y - x_2\right)\right)\expval{\mathcal{O}_{\Delta}\left(x_1\right)\mathcal{O}_{\Delta'}\left(x_2\right)}_0.
\end{equation}
Contracting Eq.~\eqref{eq:3ptCF} with $\eta_{\mu\nu}$, we can use this to fix the contact terms in that equation to
\begin{equation}
\dfrac{\eta_{\mu\nu}}{d} \, \Delta\left(\delta^{(d)}\left(y - x_1\right) + \delta^{(d)}\left(y - x_2\right)\right)\expval{\mathcal{O}_{\Delta}\left(x_1\right)\mathcal{O}_{\Delta'}\left(x_2\right)}_0.
\end{equation}
In the case of a conformally coupled free scalar, it is straightforward to check that these contact terms are consistent with those arising from the ordinary gravitational stress-energy tensor in Eq.~\eqref{eq:TcanGrav}.

Including the contact term contribution in Eq.~\eqref{eq:GenericCorrection} and running through the same steps, we find
\begin{equation}
\expval{\mathcal{O}_{\Delta}\left(x_1\right) \mathcal{O}_{\Delta'}\left(x_2\right)}_1 \supset -\frac{C_d\, \Delta}{8 d\left(\Delta x\right)^{2\Delta}} \Delta x^\mu \Delta x^\nu \partial_\mu \partial_\nu h\left(x_0\right) \, \delta_{\Delta \Delta'}\, .
\end{equation}
This simplifies the result of the first order correction to the two-point function to 
\begin{equation}
\label{eq:FirstOrderConformalTwoPoint}
\begin{aligned}
\expval{\mathcal{O}_{\Delta}\left(x_1\right) \mathcal{O}_{\Delta'}\left(x_2\right)}_1 &= -\frac{C_d \left(\Delta/\Delta_0\right)}{24 \left(d - 1\right)\left(\Delta x\right)^{2\Delta - 2}}\delta_{\Delta \Delta'} \times \\
&\hspace{0.4cm} \Bigg(R\left(x_0\right) - \frac{2\left(d - 1\right)}{(\Delta x)^2} \Delta x^\mu \Delta x^\nu R_{\mu\nu}\left(x_0\right) \\
&\hspace{0.6cm} + \frac{\left(d - 1\right)\Delta_0}{\left(\Delta x\right)^4} \Delta x^\mu \Delta x^\nu \Delta x^\rho \Delta x^\sigma \partial_\rho \partial_\sigma h_{\mu\nu}\left(x_0\right)\Bigg).
\end{aligned}
\end{equation}

We can rewrite these functions in terms of biscalar invariants, such as the geodesic distance and the van Vleck determinant, as before. The result is
\begin{equation} \label{CFT 2pt f}
\expval{\mathcal{O}_\Delta\left(x_1\right)\mathcal{O}_{\Delta'}\left(x_2\right)}_g =  \dfrac{C_d}{s^{2\Delta}}\left(\mathcal{D}_{12}^{\Delta/(2\Delta_0)}- \frac{\left(\Delta/\Delta_0\right)}{24\left(d - 1\right)} \,s^{2} \,R\left(x_0\right)\right)\delta_{\Delta \Delta'} + \dots \, ,
\end{equation}
where $s$ and $\mathcal{D}$ depend on $\left(x_1,x_2 \right)$, as defined in Appendix~\ref{a:GeometricInvariants}. The dots refer to higher order terms (in derivatives and/or curvature) that we did not compute, and possibly terms analytic in $s^2$ that vanish in the limit $s^2 \to 0$.
We can perform a consistency check by considering the conformally coupled free scalar, which corresponds to setting $\Delta = \Delta_0$. The result agrees with Eq.~\eqref{eq:nonminimalGF}, upon setting $\xi$ to the conformal coupling Eq.~\eqref{eq:conformalxi}. This confirms our previous results derived independently in the momentum space diagrammatic approach. 

For generic $\Delta$, the term proportional to the scalar curvature $R(x_0)$ has a non-analytic dependence from the geodesic squared distance $s^2$. Together with the $\mathcal{D}$-term, it therefore represents a genuine EFT prediction of QFT in curved spacetime.\footnote{We can perform a sanity check of our result by specializing to a conformally flat gravitational background. In this case, the curvature corrections can be obtained by means of a Weyl transformation of the flat space correlation function, see Appendix~\ref{a:Weyl}. Notice however that the conformally flat result is not sufficient to uniquely fix the dependence from gravitational invariants in an arbitrary background, in the absence of a complete classification of the possible bitensors appearing in such an expansion.} Notice that we never assumed the CFT itself to be weakly coupled, but only exploited the weakness of the gravitational field in a neighborhood of $x_0$, a consequence of the equivalence principle. The result therefore applies to two-point functions of strongly coupled CFTs, as well as weakly coupled ones. The only case in which the curvature contribution is analytic in $s^2$ is for $\Delta =1$, which implies the vanishing of the Hadamard tail for a conformally coupled free scalar in $d=4$. 

After the appearance of the first version of our paper, Ref.~\cite{Konechny:2026bqg} also derived this result. We checked that upon expanding the van Vleck determinant at leading order, our result agrees with theirs. Note that Ref.~\cite{Konechny:2026bqg} assumes a conformally flat metric and then argues that the same result holds for general metrics. We explicitly show that their argument indeed works for an arbitrary background at linear order in curvature.

\section{Discussion}
We have presented a framework to compute curvature corrections to the short-distance behavior of correlation functions of quantum field theories in an arbitrary curved background.
The method takes advantage of the equivalence principle to formulate the problem in terms of flat space perturbation theory, and then use field theory methods to calculate the universal non-analytic terms within the EFT regime of validity.
Curvature corrections are computed in a weak field and derivative expansion, and then expressed in terms of curvature invariants. We have done so at the leading non-trivial order for the two-point function of a free scalar field with arbitrary non-minimal coupling, of spin-1 and spin-2 fields, and of scalar primary operators in an arbitrary (potentially strongly coupled) CFT. In each case, we drew comparisons with the relevant literature, highlighting either improvements or corrections. In the Appendix we also use the same formalism to obtain a simple derivation of the conformal anomaly in two-dimensional CFTs with arbitrary central charge.

Differently from other, more familiar, applications of effective field theory, our approach is formulated in position space and computes universal non-analytic terms in the \emph{short-distance}, rather than \emph{low-momentum}, limit. 
We have provided a detailed argument for why such non-analytic contributions can be reliably computed in our approach, and checked this general expectation against well-known results of quantum field theory in curved spacetime (for free fields), or the case of conformally flat spacetimes (for CFTs).
One should keep in mind, of course, that these results are meaningful only for scales within the EFT regime of validity. That is, our results capture the short-distance behavior of correlators but only on scales larger than the UV cut-off of the theory, $r \gg \ell_{\rm UV}$. This is analogous to the conventional calculation of low momentum non-analyticities in the presence of an IR cut-off.

This work represents only a first step. The application of similar ideas to the case of higher-point correlation functions of QFTs in curved spacetime will be released in a companion paper. Another future direction is the extension of these ideas to higher orders in curvature or derivatives. The derivative and curvature expansions are independent beyond the leading order, and one can easily envisage situations in which either dominates. It would be, therefore, of great interest to extend this formalism to include these effects and distinguish the two expansions, although this comes with technical challenges. A further direction worth mentioning is that of applying these results to contexts of physical interest. Natural arenas for applications are black holes, gravitational self-force calculations and early universe cosmology, where the effect of quantum fields in curved spacetime gives rise to phenomena that are conceptually important in the former case and even observationally relevant in the latter two. We hope that our approach can help to explore some properties of interacting QFTs in these contexts, going beyond the case of free fields. 

As an example, even in a classical context, the quasinormal modes of (anti-)de Sitter and black hole spacetimes can be determined from the poles of retarded Green's functions~\cite{Leaver:1986gd, Berti:2009kk}. This suggests that our results on the short-distance analytic structure of  two-point functions could be relevant for computing high-frequency quasinormal modes. It would be interesting to explore this for interacting field theories, i.e. for nonlinear quasinormal modes~\cite{Lagos:2022otp}.

\acknowledgments
A.C.~is supported by the National Science Foundation Graduate Research Fellowship under Grant No. DGE-2036197. The work of A.P. is supported by the Huawei Young Talents Program at IHES. J.S. is supported by funds provided by the Center for Particle Cosmology at the University of Pennsylvania. This work is also partially supported by the U.S.~Department of Energy (award no.~DE-SC0011941). 

\appendix

\section{Some useful integral identities}\label{a:ImportantIdentities}

Throughout the text, we will use the following inverse Fourier transform,
\begin{align}
\label{eq:GenericI}
I_\alpha\left(x\right) &\equiv \int \frac{-i}{\left(k^2 - i\epsilon\right)^{\alpha}} e^{i k\cdot x} \frac{\dd[\n]{k}}{\left(2\pi\right)^\n},\\ \label{eq:GenericIresult}
&= \frac{\Gamma\left(\n/2 - \alpha\right)}{4^\alpha \pi^{\n/2}\Gamma\left(\alpha\right)\left(x^2 + i\epsilon\right)^{\n/2 - \alpha}} \; .
\end{align}

Note that $I_1\left(x\right)$ is simply the position space Feynman Green's function, $G_F\left(x\right)$. Note also that for large enough $\alpha$, $\alpha \ge d/2$, the integral is IR-divergent. So, in general the expression above has to be interpreted in the sense of dimensional regularization. We address the issue of IR divergences in Section~\ref{s:why}. There, we worked in a Euclidean signature for simplicity, but we can derive the result~\eqref{eq:GenericIresult} directly in Lorentzian signature using the complex Schwinger parameterization 
\be
\frac{-i}{\left(k^2 - i\epsilon\right)^{\alpha}} = \frac{i^{\alpha - 1}}{\Gamma\left(\alpha\right)} \int_{0}^{\infty} \dd{s} s^{\alpha - 1} e^{-is \left(k^2 - i\epsilon\right)}\; .
\ee
Then, upon completing the square and letting $\epsilon \rightarrow \epsilon/\left(4s^2\right)$, we have
\begin{equation}
I_\alpha\left(x\right) = \frac{i^{\alpha - 1}}{\Gamma\left(\alpha\right)} \int_{0}^{\infty} \dd{s} s^{\alpha - 1} \int \frac{\dd[d]{k}}{\left(2\pi\right)^d} e^{-is\left(k - \frac{x}{2s}\right)^2 + \frac{i}{4s}\left(x^2 + i\epsilon\right)}.
\end{equation}
It is apparent in this way of writing the integral that the $-i\epsilon$ prescription in momentum space becomes a $+i\epsilon$ in position space. Upon shifting the $k$-integral so that it is a Gaussian and Wick rotating, $k^0 = ik_d$, we find
\be \label{app:Schwinger}
I_\alpha (\Delta x^2) = \frac{i^{\alpha - d/2}}{\left(4\pi\right)^{d/2}\Gamma\left(\alpha\right)} \int_{0}^{\infty} s^{\alpha - 1 - d/2} e^{\frac{i}{4s}\left(\Delta x^2 + i\epsilon\right)}  \; . 
\ee
The $s$-integral now immediately yields the result of Eq.~\eqref{eq:GenericIresult}.

The following Feynman parameterization is also useful in position-space computations:
\begin{equation}
\label{eq:SymmetricFeynmanParameterization}
    \frac{1}{A^\alpha B^\beta} = \frac{\Gamma\left(\alpha + \beta\right)}{\Gamma\left(\alpha\right)\Gamma\left(\beta\right)}\int_{-1/2}^{1/2} \frac{\left(1/2 + u\right)^{\alpha - 1}\left(1/2 - u\right)^{\beta - 1}}{\left[\left(1/2 + u\right)A + \left(1/2 - u\right)B\right]^{\alpha + \beta}} \dd{u} \, ,
\end{equation}
where $\alpha$ and $\beta$ are arbitrary real parameters. The proof follows straightforwardly from standard identities on Gamma and Beta functions.

\section{Bi-tensors and their coincident limit expansion}
\label{a:GeometricInvariants}
When studying field theory in curved spacetime it can be convenient to express results in terms of geometric invariants, or more general covariant objects.
In the case of two-point functions, this requires the use of tensorial functions depending on two spacetime points, known as \textit{bitensors} \cite{DeWitt:1960fc}. For a review of bitensors, their properties and their use, see Ref.~\cite{Poisson:2011nh}. A bitensor is a function of points $x_1$ and $x_2$ that transforms covariantly. To keep track of the spacetime dependence of diffeomorphisms, derivatives with respect to $x_1$ and $x_2$ are assigned distinct indices: $\alpha, \beta, \dots$ for $x_1$ and $\mu, \nu, \dots$ for $x_2$. In the coincident limit $x_1 \to x_2$, bitensors reduce to ordinary tensors.

An example of a biscalar is the squared geodesic distance between two points $x_1$ and $x_2$:
\begin{equation}
s^2\left(x_1, x_2\right) = \int_{0}^{1} g_{\mu\nu}\left(z\right) \dv{z^\mu}{\lambda} \dv{z^\nu}{\lambda} \dd{\lambda},
\end{equation}
where $\lambda$ is an affine parameter for a geodesic curve between $x_1$ and $x_2$ (This means that the tangent vector $t^\mu= {\rm d}z^\mu/{\rm{d}\lambda}$ satisfies the geodesic equation $t^\nu \nabla_\nu t^\mu=0$.)
A related biscalar is Synge's world function $\sigma(x_1,x_2)$, which is defined as half our $s^2(x_1,x_2)$, see Ref.~\cite{Poisson:2011nh}.
For the purposes of this work, it is helpful to expand geometric invariants to first order in $h$ in the coincident limit. To that end, take $z = \bar{z}\left(\lambda\right) + \delta z\left(\lambda\right)$, where $\bar z(\lambda)$ is a flat space geodesic, and $g_{\mu\nu} = \eta_{\mu\nu} + h_{\mu\nu}$, and expand to first order in $h_{\mu\nu}$ and $\delta z$:
\begin{equation}
s^2\left(x_1, x_2\right) = \int_{0}^{1} \eta_{\mu\nu} \dv{\bar{z}^\mu}{\lambda} \dv{\bar{z}^\nu}{\lambda} \dd{\lambda} + \int_{0}^{1} \left(h_{\mu\nu} \dv{\bar{z}^\mu}{\lambda} \dv{\bar{z}^\nu}{\lambda} + 2\eta_{\mu\nu} \dv{\bar{z}^\mu}{\lambda} \dv{\delta z^\nu}{\lambda}\right)\dd{\lambda}.
\end{equation}
The boundary conditions are $\bar{z}\left(0\right) = x_1$, $\bar{z}\left(1\right) = x_2$, and $\delta z\left(0\right) = \delta z\left(1\right) = 0$. Since $\bar{z}^\mu$ corresponds to a flat space trajectory, we can use the boundary conditions to write
\begin{equation}
\bar{z}^\mu(\lambda) = \Delta x^\mu \lambda + x_1^\mu,
\end{equation}
where, as a reminder, $\Delta x^\mu = x_2^\mu - x_1^\mu$. As a result,
\begin{equation}
s^2\left(x_1, x_2\right) = \left(\Delta x\right)^2 + \int_{0}^{1} \left(h_{\mu\nu} (z) \Delta x^\mu \Delta x^\nu + 2\Delta x_\mu \dv{\delta z^\mu}{\tau}\right)\dd{\lambda}.
\end{equation}
Next, let's expand $h_{\mu\nu}$ to second order in $\Delta x$ around the midpoint $x_0$, Eq.~\eqref{eq:midpoint}. Since this is already of first order in $h$, it can be evaluated on $\bar{z}^\mu(\lambda)$. Noting that $\bar{z} - x_0 = \left(1/2 - \lambda\right)\Delta x^\mu$,
\begin{equation}
h_{\mu\nu}\left(\bar{z}\right) = h_{\mu\nu}\left(x_0\right) + \frac{1}{2}\left(\frac{1}{2} - \lambda\right)^2 \Delta x^\rho \Delta x^\sigma \partial_\rho \partial_\sigma h_{\mu\nu}\left(x_0\right).
\end{equation}
Moreover, $\delta z^\mu$ can be found from expanding the geodesic equation to first order and using the above $h_{\mu\nu}$,
\begin{equation}
\dv[2]{\delta z^\rho}{\lambda} + \frac{1}{2}\left(\frac{1}{2} - \lambda\right)\Delta x^\mu \Delta x^\nu \Delta x^\sigma \left(\partial_\mu \partial_\sigma h^\rho_{\hphantom{\rho}\nu} + \partial_\nu \partial_\sigma h^\rho_{\hphantom{\rho}\mu} - \partial^\rho \partial_\sigma h_{\mu\nu}\right) = 0.
\end{equation}
Integrating this twice and imposing the boundary conditions, we find
\begin{equation}
\delta z^\rho = -\frac{1}{4}\left(\frac{\lambda^2}{2} - \frac{\lambda^3}{3} - \frac{\lambda}{6}\right)\Delta x^\mu \Delta x^\nu \Delta x^\sigma\left(\partial_\mu \partial_\sigma h^\rho_{\hphantom{\rho}\nu} + \partial_\nu \partial_\sigma h^\rho_{\hphantom{\rho}\mu} - \partial^\rho \partial_\sigma h_{\mu\nu}\right).
\end{equation}
Plugging this back into our expression for the geodesic distance, we find that
\begin{equation}
\label{eq:SigmaExp}
s^2\left(x_1, x_2\right) = \Delta x^2 + h_{\mu\nu}\left(x_0\right)\Delta x^\mu \Delta x^\nu + \frac{1}{24}\Delta x^\mu \Delta x^\nu \Delta x^\rho \Delta x^\sigma \partial_\rho \partial_\sigma h_{\mu\nu}\left(x_0\right) + \dots .
\end{equation}

A second example of a biscalar is the van Vleck determinant $\mathcal{D}\left(x_1, x_2\right)$, which appears in many applications of quantum field theory in curved spacetime, such as the calculation of heat kernels, one loop approximations of the path integral, geodesic flows, and generic OPE coefficients \cite{VanVleck:1928zz, Visser:1992pz}. It is defined as 
\begin{equation}
\mathcal{D}\left(x_1, x_2\right) = \frac{1}{2}\det\left[-g^{\mu}_{\hphantom{\mu}\alpha}\left(x_2, x_1\right) \nabla^\alpha \nabla_\nu \, s^2\left(x_1, x_2\right)\right],
\end{equation}
where $g^{\mu}_{\hphantom{\mu}\alpha}\left(x_2, x_1\right)$ is a bitensor known as the parallel propagator and defined in terms of tetrad coefficients:
\begin{equation}
g^{\mu}_{\hphantom{\mu}\alpha}\left(x_2, x_1\right) = e^{\mu}_{a}(x_2) e^{a}_{\alpha}(x_1).
\end{equation}

As shown in Ref.~\cite{Poisson:2011nh}, one can express derivatives of $\sigma$ in terms of Riemann tensors. Expanding the determinant, one finds that in the coincident limit,
\begin{equation}
\label{eq:vvexpansion}
\mathcal{D}\left(x_1, x_2\right) = 1 + \frac{1}{6}R_{\mu\nu}\left(x_0\right)\Delta x^\mu \Delta x^\nu + \dots.
\end{equation}
One way to think about $\mathcal{D}$ geometrically is in terms of a congruence of geodesics. If $\mathcal{D} > 1$, geodesics undergo focusing, while if $\mathcal{D} < 1$, geodesics undergo defocusing. When $\mathcal{D} = 1$ (to all orders), we are in a flat space and geodesics remain parallel \cite{Poisson:2011nh}. 

\section{Conformal correlators from Weyl transformations}
\label{a:Weyl}
The two-point function in a conformally flat spacetime with metric
\begin{equation}
g_{\mu\nu}\left(x\right) = \Omega^2\left(x\right)\eta_{\mu\nu},
\end{equation}
is related to the flat-space two-point function by a conformal transformation of the fields. Taking the conformal weight of the scalar ${\cal O}$ to be $\Delta$,
\begin{equation}
\expval{T \mathcal{O}\left(x_1\right)\mathcal{O}\left(x_2\right)}_g = \Omega^{-\Delta}\left(x_1\right)\left(\frac{C_d}{\left(\Delta x^2 + i\epsilon\right)^{\Delta}}\right)\Omega^{-\Delta}\left(x_2\right).
\end{equation}
Since we are working in the weak gravity approximation, we can let $\Omega\left(x\right) = 1 + \omega\left(x\right)$. As a result, the first order contribution to the two-point function is
\begin{equation}
\expval{T \mathcal{O}\left(x_1\right)\mathcal{O}\left(x_2\right)}_1 = -\Delta \left(\omega\left(x_1\right) + \omega\left(x_2\right)\right) \frac{C_d}{\left(\Delta x^2 + i\epsilon\right)^{\Delta}}.
\end{equation}
Since we are interested in the coincident limit, we can expand this about $x_0$, noting that $x_1 - x_0 = \Delta x/2$ and $x_2 - x_0 = -\Delta x/2$, so that
\begin{equation}
\omega\left(x_1\right) + \omega\left(x_2\right) \simeq 2\omega\left(x_0\right) + \frac{1}{4}\Delta x^\rho \Delta x^\sigma \partial_\rho \partial_\sigma \omega\left(x_0\right).
\end{equation}
Therefore, the two-point function in a weak conformally flat background in the coincident limit is
\begin{equation}
\expval{T \mathcal{O}\left(x_1\right)\mathcal{O}\left(x_2\right)}_1 = -\frac{2\Delta C_d}{\left(\Delta x^2 + i\epsilon\right)^{\Delta}}\left(\omega\left(x_0\right) + \frac{1}{8}\Delta x^\rho \Delta x^\sigma \partial_\rho \partial_\sigma \omega\left(x_0\right)\right).
\end{equation}

Let us compare this to the result we obtained in the main text, Eq.~\eqref{eq:FirstOrderConformalTwoPoint}, where $\omega\left(x_0\right) = 0$. Since in this approximation $h_{\mu\nu} = 2\omega \, \eta_{\mu\nu}$, we find
\begin{align}
\partial_\rho \partial_\sigma h_{\mu\nu}\left(x_0\right) &= 2\eta_{\mu\nu} \partial_\rho \partial_\sigma\omega\left(x_0\right), \\
R\left(x_0\right) &= -2\left(d- 1\right)\Box\omega\left(x_0\right),\\
R_{\mu\nu}\left(x_0\right) &= -\left(d - 2\right)\partial_\mu \partial_\nu \omega\left(x_0\right) - \eta_{\mu\nu}\Box\omega\left(x_0\right).
\end{align}
Plugging this into the general expression \eqref{CFT 2pt f} and simplifying, we find the same result as above
\begin{equation}
\expval{T \mathcal{O}\left(x_1\right)\mathcal{O}\left(x_2\right)}_1 = -\frac{\Delta C_d}{4\left(\Delta x^2 + i\epsilon\right)^{\Delta}} \Delta x^\mu \Delta x^\nu \partial_\mu \partial_\nu \omega\left(x_0\right).
\end{equation}
In the  $d \to 2$ limit, the first order contribution in $\omega$ is regular for $\Delta x\to 0$ and we recover Eq.~\eqref{eq:2dimanswer}. This holds  for general metrics in $d=2$, since every metric is locally conformally flat in two dimensions.

\section{Rederiving the 2D conformal anomaly}
\label{a:anomaly}

In this appendix we provide an alternative derivation of the conformal anomaly in two-dimensional CFTs, based on the general formalism described in this work. We will work at first order in $h_{\mu\nu}$ as in the main text, which is why we will only be able to derive the anomaly in 2D, where it corresponds to a term linear in the curvature $R$. In higher (even) dimensions the anomaly appears at order $R^{d/2}$, which requires working to order $h^{d/2}$.

We start from the first-order gravitational response (we use Euclidean signature):
\begin{equation}
\expval{T_{\mu\nu}\left(x\right)}_g = \frac{1}{2}\int h_{\rho\sigma}\left(y\right) \expval{T_{\mu\nu}(x) T^{\rho\sigma}(y)}_0 \dd[2]{y},
\end{equation}
which will be enough to fix the anomaly in 2D because $R$ is the only local curvature invariant with the correct scaling dimension to contribute to $\expval{T^{\mu}_{\phantom{\mu}\mu}}_g$.
The two point function $\expval{T_{\mu\nu}(x) T^{\rho\sigma}(y)}_0$ is evaluated in flat space, so any contact term is proportional to a flat-space one-point function and vanishes. In momentum space this becomes
\begin{equation}
\expval{T_{\mu\nu}\left(q\right)}_g = \frac{1}{2}  \mathcal{G}_{\mu\nu,\rho\sigma}(q) h^{\rho\sigma}(q),
\end{equation}
where $\mathcal{G}_{\mu\nu,\rho\sigma}(q)$ is the momentum-space stress-energy tensor two-point function. Assuming diffeomorphism invariance gives the Ward identity $q^\mu \mathcal{G}_{\mu\nu,\rho\sigma}=0$, and similarly for the other indices. This forces $\mathcal{G}_{\mu\nu,\rho\sigma}(q)$ to be a symmetric transverse tensor in both  pairs of indices. In two-dimensions there is only one such structure, $\theta_{\mu\nu} \theta_{\rho\sigma}$, built from the projector $\theta_{\mu\nu}= \eta_{\mu\nu}-q_\mu q_\nu/q^2$
\footnote{In general $d$, on top of the structure displayed, there is also $\theta_{\mu (\rho} \theta_{\sigma) \nu}$. For $d=2$ these two structures coincide, and are just a complicated way to say something very simple: that, aligning for example $q^\mu$ with the time direction, transversality forces {\em all} entries of ${\cal G}_{\mu\nu, \rho\sigma}$ to vanish apart from ${\cal G}_{11,11}$. That is, in covariant form,
\be
{\cal G}_{\mu\nu, \rho\sigma} \propto \tilde q_\mu \tilde q_\nu \tilde q_\rho \tilde q_\sigma \; ,  \qquad \tilde q_\mu \equiv \epsilon_{\mu\alpha} q^\alpha \; .
\ee}. 
Scale invariance fixes the overall power of $q^2$:
\begin{equation}
\mathcal{G}_{\mu\nu,\rho\sigma}(q) = \kappa \, q^2 \theta_{\mu\nu} \theta_{\rho\sigma},
\end{equation}
where $\kappa$ is a constant that will be seen to be proportional to the central charge. Notice that $\mathcal{G}_{\mu\nu,\rho\sigma}(q) $ is non-analytic in momentum and no local counterterm (polynomial in $q$) can contribute to it. Taking the trace and using that $\theta^{\mu}_{\phantom{\mu}\mu}=1$, we obtain $\mathcal{G}^{\mu}_{\phantom{\mu}\mu,\rho\sigma}(q) = \kappa (q^2 \eta_{\rho\sigma}-q_\rho q_\sigma)$. Going to position space and writing the result in covariant form we arrive at 
\begin{equation}
\expval{T^{\mu}_{\phantom{\mu}\mu}}_g = \frac{\kappa}{2}  R \, ,
\end{equation}
where we used that $(\partial_{\rho}\partial_{\sigma} h^{\rho\sigma}- \Box h)$ is the linearized Ricci scalar.

Let us now relate $\kappa$ to the central charge. Going to position space, at non-coincident points we have
\begin{equation}
\expval{T_{\mu\nu}(x) T_{\rho\sigma}(0)}_0 = - \dfrac{\kappa}{4\pi} \partial_\mu \partial_\nu \partial_{\rho}  \partial_{\sigma} \log x^2 \, , 
\end{equation}
where we used $1/q^2 \rightarrow -(1/4\pi) \log x^2$ and $\Box \log x^2=0$. Moving to complex coordinates $z= x^1 + i\,  x^2$, $\bar{z}= x^1 - i\,  x^2$ and $x^2 = z \bar{z}$, and using $\partial_z^4 \log z = -6/z^4$, we arrive at 
\begin{equation}
\expval{T_{zz}(z) T_{zz}(0)} = \dfrac{3\kappa}{2\pi z^4}   \, . 
\end{equation}
Using the standard 2D CFT normalization of the stress tensor~\cite{Polchinski:1998rq}, $T= -2\pi T_{zz}$, we obtain
\begin{equation}
\expval{T(z) T(0)} = \dfrac{6\pi\kappa}{z^4} = \dfrac{c}{2  z^4} \quad \implies \quad \kappa = \dfrac{c}{12\pi}  \, . 
\end{equation}
Using the mode expansion $T(z)= \sum L_n z^{-n-2}$, the constant $c$ can be seen to correspond to the central charge of the Virasoro algebra $[L_m, L_n]= (m-n)L_{m+n}+ (c/12) m (m^2-1) \delta_{m+n,0} $.

We therefore arrive at the final result
\begin{equation}
\expval{T^{\mu}_{\phantom{\mu}\mu}}_g = \frac{c}{24 \pi}  R \, ,
\end{equation}
This derivation is quite simple and agrees with the well-known result, see for instance Ref.~\cite{DiFrancesco:1997nk}.
A related derivation for 2D CFTs, but in position space, has appeared in Ref.~\cite{Bonora:2015nqa}, where it was carried out without assuming diffeomorphism invariance (covering also chiral theories with a gravitational anomaly, $c_L\neq c_R$). See also Ref.~\cite{Bzowski:2017poo} for a different perspective on conformal anomalies in momentum space.

\bibliographystyle{JHEP}
\bibliography{biblio.bib}

\end{document}